\documentstyle[preprint,aps,epsfig]{revtex}
\draft
\title{Neutrino Reactions on Deuteron}
\author{
  S. Nakamura${}^1$, T. Sato${}^{1,2}$, V. Gudkov${}^2$
  and K. Kubodera${}^2$\\ 
  ${}^1${\it Department of Physics, Osaka University, Toyonaka, Osaka
  560-0043, Japan}\\ 
${}^2${\it Department of Physics and Astronomy, University of South Carolina,
Columbia, SC 29208, USA}} 
\begin{document}
\maketitle

\vspace{1.5cm}
\begin{center} 
ABSTRACT
\end{center}

\vspace{2mm}
The cross sections for the $\nu-d$ and 
$\bar{\nu}-d$ reactions are calculated 
for the incident energy up to $E_\nu = 170$ MeV,
with the use of a phenomenological Lagrangian approach.
We assess and improve the reliability
of the employed calculational method
by examining the dependence of the results
on various input and approximations
that go into the calculation.
The main points of improvements
over the existing work are:
(1) use of the ``modern" NN potentials;
(2) use of the more accurate nucleon weak-interaction
form factors;
(3) monitoring the strength of a vertex that governs 
the exchange-current contribution,
with the use of data on the related process,
$n+p\rightarrow d+\gamma$.
In addition to the total cross sections,
we present various differential cross sections
that are expected to be useful 
for the SNO and other experiments.
In the low energy regime relevant to
the solar neutrinos,
the newly calculated total cross sections
essentially agree with the existing literature values.
The origins of slight differences found for higher energies
are discussed.
The ratio between the neutral-current 
and charged-current reaction cross sections
is found to be extremely stable 
against any variations in the input 
of our calculation.

\newpage

\section{Introduction}
The neutrino-deuterium reactions\footnote{
When convenient, we use the word ``neutrino" and  
the symbol ``$\nu$" in a generic sense,
referring to both neutrinos and anti-neutrinos.}
have been studied extensively over the past decades 
\cite{ellis,PRD12-3673,NPA294_473,avi,nozetal,bkn,YHH1,TKK,DK1,YHH2,kk92,kubo1,EFT,EFT2}.
Recent detailed studies are strongly motivated
by the proposal and successful start
of the Sudbury Neutrino Observatory 
(SNO) \cite{SNO},
which uses a large underground heavy-water 
Cerenkov counter.
One of the primary goals of SNO
is to study the solar neutrinos by monitoring
three reactions occurring in heavy water:
(i) $\nu-e$ scattering,
$\nu_e + e^-\rightarrow  \nu_e + e^-$;
(ii) the charged-current (CC) reaction,
$\nu_e + d \rightarrow  e^- + p + p$;
(iii) the neutral-current (NC) reaction,
$\nu_x + d \rightarrow  \nu_x + n + p$,
where $\nu_x$ stands for a neutrino of any flavor. 
The unique feature of SNO is its ability 
to register the CC and NC reactions 
separately but simultaneously.
Since the NC reaction measures the total flux 
of the solar neutrinos (regardless of their flavors),
SNO experiments offer valuable information
about the nature of possible neutrino oscillation.
SNO is also capable of monitoring astrophysical neutrinos
the energy of which extends
well beyond the solar neutrinos energy regime,
a prominent example being supernova neutrinos.
Obviously, in interpreting experimental results 
to be obtained at SNO,
the accurate knowledge of the $\nu-d$ reaction 
cross sections is a prerequisite.
Although the $\nu-e$ scattering cross section
is readily available from the standard model,
estimation of the neutrino-deuteron reaction 
cross sections requires a detailed examination 
of the structure of two-nucleon systems
and their responses to electroweak probes.

In describing the current theoretical situation
regarding the $\nu-d$ cross sections,
it is useful to consider the $\nu-d$ reactions
in a broader context of the general responses
of two-nucleon systems to electroweak probes.
A highly successful method for describing
these responses is to consider one-body
impulse approximation terms and
two-body exchange-current terms
acting on non-relativistic nuclear wave functions,
with the exchange currents derived
from a one-boson exchange model.
In a modern realization of this approach
\cite{cr,it77,towner},
the vertices characterizing relevant Feynman diagrams
are determined, as much as possible,
with the use of the low-energy theorems
and current algebra.
Some coupling constants are inferred from 
models (the quark model, SU(3), SU(6), etc.).
In the present work we refer to this type
of formalism as the 
phenomenological Lagrangian approach (PhLA).
This formalism has been used extensively 
for electromagnetic processes in
two nucleon systems \cite{riska-brown,fm,n-cap}.
The reported good agreement between theory and experiment
gives a strong hint of the basic soundness of PhLA.
This method has also been applied to 
two-nucleon weak-interaction processes
such as muon capture on the deuteron
\cite{TKK,drr,Doi-muon}, 
the $pp$-fusion reaction \cite{drr,crsw},
and the $\nu-d$ reactions.
For the muon capture,
the calculated capture rate agrees reasonably well
with the experimental value,
again rendering support for the basic legitimacy 
of PhLA.
(For the $pp$ fusion there is unfortunately no data 
available.)

For the neutrino-deuterium reactions,
the most detailed study within the framework
of the impulse approximation (IA) 
has been done by Ying, Henley and Haxton 
(YHH) \cite{YHH2},
while the most elaborate PhLA calculations
including the exchange-current effects as well
as the IA terms have been carried out
in \cite{TKK,DK1,YHH2,kk92},
and the latest status is described in \cite{kubo1}, 
to be referred to as KN.\footnote{
Ref.\cite{kubo1} also gives a rather detailed account
of the relation between these latest calculations
and the earlier work.}
In the solar neutrino energy regime,
the cross sections given in KN are slightly larger 
than those of YHH.
This difference, however, is mostly due 
to the absence of the exchange-current 
contributions in YHH.
As far as comparison with data is concerned,
Tatara {\it et al.}'s estimate \cite{TKK} of 
$\sigma(\nu_e+d\rightarrow e^-+p+p)$ 
averaged over
the Michel spectrum of $\nu_e$ agrees 
with the result of a stopped-pion-beam experiment
\cite{Willis} within large experimental errors
(30\%).
Furthermore, the result of 
a Bugey reactor neutrino experiment 
\cite{bugey} agrees, 
within 10\% experimental errors, with 
the values of 
$\sigma(\bar{\nu}_e+d\rightarrow e^++n+n)$ and 
$\sigma(\bar{\nu}_e+d\rightarrow
\bar{\nu}_e+p+n)$ 
given in  KN.
Thus PhLA seems to provide a reasonably reliable
framework for calculating
the neutrino-deuteron cross sections.

Meanwhile, a new approach based on effective field theory
(EFT) has been scoring great success in describing 
low-energy electroweak processes in the
two-nucleon systems \cite{pmr,pkmr,kol,ksw,pds2,pt99}.
In particular, the rate of thermal neutron radiative 
capture on the proton 
($n+p\rightarrow d+\gamma$)
has been calculated in chiral perturbation theory
($\chi$PT) and the result is found to be 
in perfect agreement with the data \cite{pmr}.
Butler and Chen \cite{EFT} and 
Butler, Chen and Kong \cite{EFT2} have recently  made
extremely elaborate studies of $\nu-d$ cross sections
for the solar neutrino energies with the use of EFT.
The results of their EFT calculation
agree with those of PhLA in the following sense.
In an EFT approach, one starts with a general effective
Lagrangian, ${\cal{L}}_{eff}$, 
that contains all possible terms compatible 
with given symmetries and a given order of expansion;
the coefficient of each term in ${\cal{L}}_{eff}$
is called the low-energy coefficient (LEC).
Now, it often happens that some LEC's cannot be fixed 
by the symmetry requirements alone and hence 
need to be treated as parameters to be determined
empirically. 
In \cite{EFT,EFT2}, 
the coefficient $L_{\rm 1A}$ of
a four-nucleon axial-current counter term
enters as an unknown parameter,
although dimensional arguments suggest
$-6\,{\rm fm}^3\le L_{\rm 1A}\le +6\,{\rm fm}^3$.
According to \cite{EFT},
the $\bar{\nu}-d$ cross sections obtained in EFT
agree with those of the PhLA calculation
(YHH or KN),
provided $L_{\rm 1A}$ is adjusted appropriately.
The optimal value of $L_{\rm 1A}$ is
$L_{\rm 1A}=6.3\,{\rm fm}^3$ for YHH,
and 
$L_{\rm 1A}=1.0\,{\rm fm}^3$ for KN,
reasonable values as compared 
with the above-mentioned dimensional estimates. 
The fact that an {\it ab initio} 
calculation (modulo one free parameter)
based on EFT is consistent with the results of PhLA
provides further evidence for the basic reliability 
of PhLA.

Bahcall, Krastev and Smirnov 
\cite{bks00} have recently studied in great detail
the consequences of measurements 
of various observables at SNO.
As input for their analysis,
the $\nu_e-d$ reaction cross sections
of YHH and KN are used,
and the difference between these two calculations
are assumed to represent 1$\sigma$ theoretical errors.
According to \cite{bks00},
uncertainties in the $\nu-d$ cross sections
represent the largest ambiguity
in most physics conclusions obtainable
from the SNO observables,
a feature that again points to the importance of reducing 
the uncertainty in the $\nu-d$ reaction cross sections.

In the present article we carry out,
within the framework of PhLA, 
a detailed study of the cross sections
for the charged-current (CC)
and neutral-current (NC) reactions of neutrinos 
and anti-neutrinos with the deuteron:
\begin{eqnarray}
  \nu_e+d & \rightarrow & 
  e^- + p + p, \label{nuCC}\\
  \nu_x + d & \rightarrow & 
  \nu_x + p + n \;\;\;\;\;\;\;\;\;
  (x=e,\,\mu,\,{\rm or}\,\,\tau), \label{nuNC}\\
  \bar{\nu}_e + d & \rightarrow &
   e^+ + n + n, \label{nubarCC}\\
  \bar{\nu}_x + d & \rightarrow & 
  \bar{\nu}_x + p + n \;\;\;\;\;\;\;\;\;
  (x=e,\,\mu,\,{\rm or}\,\,\tau)\,.\label{nubarNC}
\end{eqnarray}
It is our view that, in calculating the low-energy 
$\nu-d$ cross sections, EFT and PhLA play 
complementary roles.
EFT, being a general framework, is capable of giving 
model-independent results, \underbar{\it provided}
all the LEC's in an effective Lagrangian 
${\cal{L}}_{eff}$ are predetermined. 
At present, however, ${\cal{L}}_{eff}$ does 
contain an unknown LEC, $L_{\rm 1A}$.\footnote{
In principle, however, it is possible to fix  
$L_{\rm 1A}$ using a parity-violating
electron scattering experiment.\cite{EFT,EFT2}}
Meanwhile, although PhLA is a model approach,
its basic idea and the parameters contained in it
have been tested using many observables.
Thus, insofar as one accepts the validity of 
these tests, PhLA has predictive power.
It is reassuring that, as mentioned,
there is highly quantitative correspondence \cite{EFT,EFT2}
between the low-energy $\nu-d$ cross sections 
obtained in PhLA and those of EFT within 
a reasonable range for $L_{\rm 1A}$.
In this article we wish to investigate several key aspects 
of PhLA in more depth than hitherto reported.  
 
Beyond the solar neutrino energy regime,
PhLA is at present the only available formalism
for evaluating the $\nu-d$ cross sections.
The EFT calculation in \cite{EFT,EFT2},
by design, ``integrates out" all the degrees of freedom
but that of the heavy baryon.
The nature of this so-called ``nucleon-only" EFT 
limits its applicability to very low incident neutrino energies 
(typically the solar neutrino energies).\footnote{
One can hope to extend the applicability of EFT to higher
energies by including the pion degree of freedom explicitly
via $\chi$PT.  An {\it ab initio} calculation based on $\chi$PT
for the $\nu-d$ reactions is yet to be done.}
On the other hand, there is no obvious conceptual obstacle
in using PhLA in an energy regime significantly higher
than that of the solar neutrinos.
Therefore, once the reliability of PhLA is tested
at low energies by comparison with experimental data
or with the results of EFT, 
it is rather natural to use PhLA for higher
energies as well.
In this sense, too, EFT and PhLA 
seem to play complementary roles (at least,
in the current status of the matter).

Our main goals here is to assess and improve the reliability
of the PhLA calculation of the $\nu-d$ reaction cross sections,
by carefully examining the dependence of the results
on various input and approximations
that go into calculations.
The main points of improvements
in this work over the existing estimates are:
(1) Use of the ``modern" NN potentials;
(2) Use of the more accurate nucleon weak-interaction
form factors;
(3) Monitoring the strength of the 
$\pi N\Delta$ vertex that governs 
by far the dominant exchange-current contribution,
with the use of data on the related process,
$n+p\rightarrow d+\gamma$.
A second practical goal of this paper
is to provide detailed information 
about the various differential cross sections
for the $\nu-d$ reactions.
Although the total cross sections are well documented 
in the literature,
there have not been systematic descriptions 
of the differential cross sections.
We therefore discuss in detail
the energy spectrum, angular distribution 
and double differential cross sections
of the final lepton in the CC reaction 
and also the energy spectrum and angular distribution
of the final neutron in the NC reaction.
It is hoped that, the detailed information
given here on these differential cross sections
will be useful in analyzing SNO and other experiments.

In the low energy regime relevant to
the solar neutrinos,
our results are found to be in essential agreement
with those of KN.
Based on these and additional results
described in this article, 
we shall deduce the best estimates
of theoretical errors
in the $\nu-d$ cross sections.
For higher energies, the present calculation
gives $\nu-d$ total cross sections larger
than those of KN by up to 6\%; we shall discuss
the origin of this variance.

The organization of the rest of this article is as follows.
After giving in Sec. II a brief account 
of the general framework of our PhLA,
we describe in Sec. III the calculational details,
including the multipole expansion 
of the nuclear currents,
and expressions for the cross sections
for $\nu-d$ reactions. 
The numerical results are presented in Sec. IV,
and discussion and summary are given in Sec. V.
Some kinematical formulae necessary
for calculating phase space integrals are given
in Appendix.

\section{Formalism} \label{section_form}

We are concerned with the $\nu/\bar{\nu}-d$ reactions 
listed in Eq.(\ref{nuCC})-Eq.(\ref{nubarNC}).
The four-momenta of the participating particles 
are labeled as
\begin{equation}
\nu/\bar{\nu}(k)+d(P) \rightarrow \ell(k')+N_1(p_1')+N_2(p_2'),
\label{momdef}
\end{equation}
where $\ell$ corresponds to $e^{\pm}$ for the CC reactions
\ [Eqs.(\ref{nuCC}),(\ref{nubarCC})], 
and to $\nu$ or $\bar{\nu}$ for the NC reactions
\ [Eqs.(\ref{nuNC}),(\ref{nubarNC})].
The energy-momentum conservation reads
$k+P=k'+P'$ with $P'\equiv p_1'+p_2'$,
and we denote a momentum transfer from
lepton to nucleus by 
$q^{\mu} = k^{\mu} - k'^{\mu} = P'^{\mu} - P^{\mu}$.
In the laboratory system to be used throughout this work, 
we write
\begin{equation}
k^\mu=(E_{\nu},\bbox{k}),\,\,
k'^\mu=(E'_{\ell},\bbox{k}'),\,\,
P^\mu=(M_d,\bbox{0}),\,\,
P'^\mu=(P'^0,\bbox{P}'),\,\,
q^\mu=(\omega,\bbox{q}).
\end{equation}

The interaction Hamiltonian 
for semileptonic weak processes is
given by the product of 
the hadron current ($J_{\lambda}$) and 
the lepton current ($L^{\lambda}$) as\footnote{
Throughout we use the Bjorken and Drell convention
for the metric and Dirac matrices, except that we adopt
the Dirac spinor normalized as $u^{\dagger}u = 1$.
}
\begin{eqnarray}
H_W^{CC} & = & \frac{G_F \cos \theta_C}
{\sqrt{2}}\int d\bbox{x}  [
   J_{\lambda}^{CC}(\bbox{x})L^{\lambda}(\bbox{x}) +
    \mbox{h. c.}]\label{eq_Ham-CC}
\end{eqnarray}
for the CC process and
\begin{eqnarray}
H_W^{NC} & = & \frac{G_F}{\sqrt{2}}\int d\bbox{x}
\  [J_{\lambda}^{NC}(\bbox{x})L^{\lambda}
    (\bbox{x})+\mbox{h. c.}]\label{eq_Ham-NC}
\end{eqnarray}
for the NC process.
Here $G_F = 1.166 \times 10^{-5}\;\mbox{GeV}^{-2}$
is the Fermi coupling constant,
and $\cos \theta_C=0.9749$ is the Cabibbo angle.

The lepton current is given by
\begin{eqnarray}
L^{\lambda}(\bbox{x}) & = & \bar{\psi}_{l}(\bbox{x})
\gamma^{\lambda}(1-\gamma^5)\psi_{\nu}(\bbox{x}),
\end{eqnarray}
and its matrix element is written as 
\begin{eqnarray}
   l^{\lambda} \equiv
<k'|L^{\lambda}(0)|k> 
&=& \bar{u}_{l}(k')\gamma^{\lambda}
(1-\gamma^5)u_{\nu}(k)\;\;\;\;\;\;\;\;
{\rm for\,\,} \nu{\rm -reaction}\,,\nonumber \\
&=& \bar{v}_{\bar{\nu}}(k) 
\gamma^{\lambda}(1-\gamma^5)v_{\bar{l}} (k')
\;\;\;\;\;\;\;\;{\rm for\,\,} \bar{\nu}{\rm -reaction}\,.\label{eq_l-def}
\end{eqnarray}

The hadronic charged current has the form
\begin{eqnarray}
J_{\lambda}^{CC}(\bbox{x}) & = & 
V_{\lambda}^{\pm}(\bbox{x}) +
A_{\lambda}^{\pm}(\bbox{x}) ,
\end{eqnarray}
where $V_{\lambda}$ and $A_{\lambda}$ 
denote the vector and axial-vector currents, respectively.
The superscript $+(-)$  denotes 
the isospin raising (lowering) operator  for
the $\bar{\nu}(\nu)$-reaction.
Meanwhile, according to the standard model,
the hadronic neutral current is given by
\begin{eqnarray}
J_{\lambda}^{NC}(\bbox{x}) 
&=& (1-2 \sin^2 \theta_W )V_{\lambda}^{3} +
A_{\lambda}^{3} -2 \sin^2 \theta_W 
V_{\lambda}^{s} , \label{eq_NC-current}
\end{eqnarray}
where $\theta_W$ is the Weinberg angle 
with $\sin^2 \theta_W = 0.2312$. 
$V_{\lambda}^{s}$ is the isoscalar part 
of the vector current,
and the superscript `3' denotes 
the third component of the isovector current. 
In the present case
the hadron current consists of 
one-nucleon impulse approximation (IA) terms
and two-body meson exchange current (MEX) terms.
Their explicit forms are described 
in the next subsections.

\subsection{Impulse approximation current}

The IA current is determined by the single-nucleon
matrix elements of $J_\lambda$.
The nucleon matrix elements of the currents are
written as 
\begin{eqnarray}
<\!N(p')\ |\ V_{\lambda}^{\pm}(0)\ |\ N(p)\!> & = &
  \bar{u}(p')[f_V \gamma_{\lambda} + i \frac{f_M}{2M_N}
    \sigma_{\lambda\rho}q^\rho]\tau^{\pm} u(p), \\
<\!N(p')\ |\ A_{\lambda}^{\pm}(0)\ |\ N(p)\!> & = &
  \bar{u}(p')[f_A \gamma_{\lambda}\gamma^5 +
  f_P \gamma^5 q_{\lambda} ] \tau^{\pm} u(p)\,,
\end{eqnarray}
where $M_N$ is the average of the masses 
of the final two nucleons.
For the third component of the isovector current, 
we simply replace $\tau^{\pm}$ with $\frac{\tau^3}{2}$. 
For the isoscalar current 
\begin{eqnarray}
<\!N(p')\ |\ V_{\lambda}^{s}(0)\ |\ N(p)\!> & = &
  \bar{u}(p')
[f_V \gamma_{\lambda} + i \frac{f_M^s}{2M_N}
  \sigma_{\lambda\rho}q^\rho]\frac{1}{2} u(p).
\end{eqnarray}
The non-relativistic forms of the IA currents are given by
\begin{eqnarray}
V_{IA,0}^{\pm}(\bbox{x}) & = & 
\sum_i f_V \tau_i^{\pm}\delta(\bbox{x}-\bbox{r}_i), \\
\bbox{V}_{IA}^{\pm}(\bbox{x}) & = & 
\sum_i [f_V \frac{\bbox{p}_i'+\bbox{p}_i}{2M_N}
       +  \frac{f_V + f_M}{2M_N}
\bbox{\nabla}\times\bbox{\sigma}_i ]
          \tau_i^{\pm}\delta(\bbox{x}-\bbox{r}_i), \\
A_{IA,0}^{\pm}(\bbox{x}) & = & \sum_i [ \frac{f_A}{2M_N}
\bbox{\sigma}_i\cdot(\bbox{p}_i'+\bbox{p}_i)
       - \frac{i f_P \omega}{2M_N} 
\bbox{\sigma}_i\cdot\bbox{\nabla} ]
          \tau_i^{\pm}\delta(\bbox{x}-\bbox{r}_i), \\
\bbox{A}_{IA}^{\pm}(\bbox{x}) & = & 
\sum_i[f_A \bbox{\sigma}_i +
\frac{f_P}{2M_N}\bbox{\nabla}( \bbox{\nabla}\cdot 
\bbox{\sigma}_i) ]
\tau_i^{\pm}\delta(\bbox{x}-\bbox{r}_i),\\
V_{IA,0}^{s}(\bbox{x}) & = & \sum_i f_V\ 
\frac{1}{2}\ \delta(\bbox{x}-\bbox{r}_i),\\
\bbox{V}_{IA}^s(\bbox{x}) & = & 
\sum_i [f_V \frac{\bbox{p}_i'+\bbox{p}_i}{2M_N}
       +  \frac{f_V + f_M^s}{2M_N}
\bbox{\nabla}\times\bbox{\sigma}_i ]
         \ \frac{1}{2}\ \delta(\bbox{x}-\bbox{r}_i).
\end{eqnarray}
It is useful to rewrite 
$\bbox{p}_i + \bbox{p}_i'= \bbox{q} + 
\bbox{P} \pm 2\bbox{p}_N$,
where the $+ (-)$ sign corresponds to $i=1$ ($i=2$),
and the derivative operator $\bbox{p}_N$ should act
on the deuteron wave function;
in the laboratory system we are working in, 
we have $\bbox{P} = \bbox{0}$.

As for the $q_{\mu}^2$ dependence of the form factors
we use the results of the latest analyses in
\cite{cvm,avm}:
\begin{eqnarray}
  f_V(q_{\mu}^2) &=& G_D(q^2_\mu)(1 + \mu_p \eta)(1 + \eta)^{-1},
  \label{eq:fv}\\
  f_M(q_{\mu}^2) &=& G_D(q^2_\mu)
   (\mu_p - \mu_n - 1 - \mu_n \eta)(1 + \eta)^{-1},\\ 
  f_A(q_{\mu}^2) &=& -1.254\ G_A(q^2_\mu),\\
  f_P(q_{\mu}^2) &=& \frac{2M_N}{m^2_{\pi}-q_{\mu}^2}f_A(q_{\mu}^2),\\
  f_M^s(q_{\mu}^2) &=& G_D(q^2_\mu)
   (\mu_p + \mu_n - 1 + \mu_n \eta)(1 + \eta)^{-1},
   \label{eq:fsM}
\end{eqnarray}
with
\begin{eqnarray}
G_D(q^2_\mu) & = & \left(1 -  \frac{q^2_\mu}
{0.71\mbox{GeV}^2}\right)^{-2},\\
G_A(q^2_\mu) & = & \left(1 -  \frac{q^2_\mu}
{1.14\mbox{GeV}^2}\right)^{-2},
\label{eq:GA}
\end{eqnarray}
where $\mu_p=2.793$, $\mu_n = -1.913$, $\eta = -
\frac{q_\mu^2}{4M_N^2}$ and $m_{\pi}$ is the pion mass.

\subsection{Exchange currents}

As mentioned, we use a phenomenological Lagrangian 
approach (PhLA) to estimate the contributions
of meson-exchange (MEX) currents.
In a PhLA due to Ivanov and Truhlik \cite{it77},
the MEX operators are derived in a hard pion approach
\cite{oz70}, in which one explicitly constructs
a phenomenological Lagrangian consistent with current 
algebra, PCAC, and the vector meson dominance.
This Lagrangian was used 
by Tatara {\it et al.\ }\cite{TKK}
in their calculations for $\mu-d$ capture
and the $\nu-d$ reactions \cite{TKK}.
Meanwhile, the studies by Doi {\it et al.\ }
\cite{DK1,Doi-muon} indicate that only 
a small subset of the possible diagrams 
gives essentially the same results as the full set.
Based on this experience, we consider here 
the following types of exchange currents.
 
\vspace{3mm}
\noindent
{\bf Axial-vector current}

The axial vector exchange current, $A^{\mu}_{MEX}$, 
consists of a pion-pole term and a non-pole part,
$\bar{A}^{\mu}_{MEX}$. 
Using the PCAC hypothesis, 
we can express $A^{\mu}_{MEX}$ in terms of 
the non-pole part alone:
\begin{eqnarray}
  A^{\mu}_{MEX} = \bar{A}^{\mu}_{MEX} -
  \frac{q^{\mu}}{m^2_{\pi}-q_{\mu}^2} ( \bbox{q}\cdot
  \bbox{\bar{A}}_{MEX} - \omega \bar{A}_{MEX,0} ).
\end{eqnarray}
We therefore need only consider the non-pole part.
For the time component it is known that one-pion exchange
diagram gives the most important contribution,
called the KDR current \cite{KDR}.\footnote{
As discussed extensively in \cite{krt92,tow92},
corrections to the KDR current can arise from 
heavy-meson exchange diagrams.
We however do not consider those corrections here,
since the contribution of the KDR current
in the present case turns out to be small
(see below).}
The explicit form of the KDR current,
with a vertex form factor supplemented, reads
\begin{eqnarray}
\bar{A}^{\pm}_{KDR,0} (\bbox{x}) &=& \frac{1}{if_A}
  \Biggl(\frac{f}{m_\pi}\Biggr)^2 
  \delta(\bbox{x} - \bbox{r}_1)
[\bbox{\tau}_1 \times
  \bbox{\tau}_2]^{(\pm)} \int 
\frac{d \bbox{q'}}{(2 \pi)^3}
  K^2_\pi(\bbox{q'}^2) 
\frac{e^{-i \bbox{q'}\cdot
      \bbox{r}}}{\omega^2_\pi}
(\bbox{\sigma}_2 \cdot \bbox{q'})
  \nonumber \\
& &  + (1 \leftrightarrow 2)\label{eq_KDR},
\end{eqnarray}
with $\bbox{r}=\bbox{r}_1-\bbox{r}_2$ and
$\omega_\pi=\sqrt{\bbox{q'}^2+m_\pi^2}$.
For the space component, 
we take account of the isobar current,
$\bbox{\bar{A}}^{\pm}_{\Delta}$,
that arises from one-pion and
one-$\rho$-meson exchange diagrams. 
Its explicit form is  
\begin{eqnarray}
  \bbox{\bar{A}}^{\pm}_{\Delta}(\bbox{x}) &=& 
4 \pi f_A \delta(\bbox{x}-\bbox{r}_1) 
  \int\frac{d \bbox{q'}e^{-i \bbox{q'}\cdot 
\bbox{r}}}{(2 \pi)^3} \nonumber \\
& & \times \Biggl[ 
\frac{K^2_\pi(\bbox{q'}^2)}
{\omega^2_\pi} \{ c_0 \bbox{q'}
  \bbox{\tau}_2^{(\pm)}+d_1(\bbox{\sigma}_1 
  \times \bbox{q'})[\bbox{\tau}_1
  \times \bbox{\tau}_2]^{(\pm)} \} 
(\bbox{\sigma}_2 \cdot \bbox{q'})
  \nonumber \\
& & +\frac{K^2_\rho(\bbox{q'}^2)}
{\omega^2_\rho} \left\{c_ \rho \bbox{q'} \times
  (\bbox{\sigma}_2 \times \bbox{q'})  
\bbox{\tau}_2^{(\pm)}+d_ \rho
  \bbox{\sigma}_1 \times  (\bbox{q'} \times 
  (\bbox{\sigma}_2 \times \bbox{q'}))[\bbox{\tau}_1
  \times \bbox{\tau}_2]^{(\pm)}\right\} \Biggr] 
\nonumber \\
& & + (1 \leftrightarrow 2)\label{eq_delta},
\end{eqnarray}
with $\omega_\rho=\sqrt{\bbox{q'}^2+m_\rho^2}$
and $m_{\rho}$ is the mass of $\rho$-meson.
For the third component
of the isovector current, we just replace $\tau_i^{\pm}$ and 
$[\bbox{\tau}_1\times\bbox{\tau}_2]^{(\pm)}$ with 
$\frac{\tau_i^3}{2}$ and 
$\frac{[\bbox{\tau}_1 \times\bbox{\tau}_2]^{(3)}}{2}$, respectively. 
(This prescription will be applied 
to other exchange currents as well.) 
The numerical values of the pion coupling constants 
can be determined from low-energy
pion-nucleon scattering \cite{isobar}, 
while the $\rho$-meson coupling constants
are deduced from the quark model:
\begin{eqnarray}
\frac{f^2}{4\pi}=0.08,\quad c_0m^3_\pi = 
0.188,\quad d_1m^3_\pi = -0.044,\quad
c_ \rho m^3_ \rho = 36.2,\quad  d_ \rho = 
- \textstyle{\frac{1}{4}}c_ \rho . \nonumber
\end{eqnarray}
Furthermore, we assume that the $\bbox{q}^2$-dependence 
of the vertex form factors, $K_{mN}(\bbox{q}^2)$ 
and $K_{m \Delta}(\bbox{q}^2)$ $(m= \pi,\rho)$,
are given by
$  K_{\pi N}(\bbox{q}^2)=K_{\pi  \Delta}(\bbox{q}^2)
  =K_{\pi}(\bbox{q}^2)=
  \frac{\Lambda^2_{\pi}-m^2_{\pi}}
{\Lambda^2_{\pi}+\bbox{q}^2} ,
  K_{\rho N}(\bbox{q}^2)=K_{\rho  \Delta}(\bbox{q}^2)
  =K_{\rho}(\bbox{q}^2)=
  \frac{\Lambda^2_{\rho}-m^2_{\rho}}
  {\Lambda^2_{\rho}+\bbox{q}^2}$,
with cutoff masses, $\Lambda_{\pi}$ = 
1.18 GeV and $\Lambda_{\rho}$ = 1.45 GeV \cite{cutoff}.  
We use the above-listed values of 
coupling constants and form factors 
as our standard parameters.

\vspace{5mm}
\noindent
{\bf Vector current}

Regarding the vector exchange currents, 
we first note that the exchange currents 
for the time component must be small, 
since the exchange currents for charge 
vanish in the static limit.
As for the space component,
we take into account pair, pionic, and
isobar currents.
If we adopt the one-pion exchange model 
for the pair and pionic current and 
the one-pion and one-$\rho$-meson exchange model 
for the isobar current, 
their explicit forms are given as
\begin{eqnarray}
  \bbox{V}^{\pm}_{pair}(\bbox{x}) &=& 
-2if_V \left( \frac{f}{m_ \pi} \right)^2
  \delta(\bbox{x}-\bbox{r}_1)
[\bbox{\tau}_1 \times \bbox{\tau}_2]^{(\pm)}
\int \frac{d \bbox{q'}}{(2 \pi)^3}K^2_\pi(\bbox{q'}^2) 
\frac{e^{-i \bbox{q'}\cdot
      \bbox{r}}}{\omega^2_\pi} 
\bbox{\sigma}_1 (\bbox{\sigma}_2 \cdot \bbox{q'})
  \nonumber \\
& &  + (1 \leftrightarrow 2)\,,\label{eq_pair}
\end{eqnarray}
\begin{eqnarray}
\bbox{V}^{\pm}_{pionic}(\bbox{x}) &=& 2i 
\left( \frac{f}{m_ \pi} \right)^2
    [\bbox{\tau}_1 \times \bbox{\tau}_2]^{(\pm)}
\int \frac{d \bbox{q'}_1}{(2 \pi)^3}K_\pi(\bbox{q'}^2_1)
\int \frac{d \bbox{q'}_2}{(2 \pi)^3}K_\pi(\bbox{q'}^2_2) 
\nonumber\\
& &\times\frac{e^{-i
  \bbox{q'}_1\cdot(\bbox{r}_1-\bbox{x})}}
{\omega^2_{\pi 1}}
  \frac{e^{-i \bbox{q'}_2\cdot 
(\bbox{x}-\bbox{r}_2)}}{\omega^2_{\pi 2}}
 (\bbox{\sigma}_1 \cdot 
\bbox{q'}_1)(\bbox{\sigma}_2 \cdot
\bbox{q'}_2)(\bbox{q'}_1+\bbox{q'}_2 ),\\
\bbox{V}^{\pm}_{\Delta}(\bbox{x}) &=& 
- \frac{f_V + f_M}{2 M_N f_A}\bbox{\nabla} \times
\bbox{\bar{A}}^{\pm}_{\Delta},
\end{eqnarray}
with $\omega_{\pi i}=\sqrt{m_\pi^2 + \bbox{q}_i^{'2}}$.

\subsection{Nucleon-nucleon potential}

In PhLA, the nuclear transition matrix elements 
are obtained by sandwiching the one-body IA 
and two-body MEX currents 
between the initial and final nuclear wave functions
which obey the Schr\"{o}dinger equation
that involves a phenomenological nucleon-nucleon potential.
The earlier work \cite{TKK,DK1} indicates that,
so long as we use a realistic NN potential
that reproduces with sufficient accuracy
the scattering phase shifts
and the deuteron properties, the numerical results for 
the $\nu-d$ cross sections are not too
sensitive to particular choices of NN potentials.
It seems worthwhile to further check this stability
for the {\it modern} potentials that were not
available at the time of the  work described in \cite{TKK,DK1}.
As representatives of the ``state-of-the-art"
NN potentials, we consider in this work the following three:
the Argonne-$v18$ potential (ANLV18) \cite{anlv18},
the Reid93 potential \cite{nij-Reid93},
and the Nijmegen II potential (NIJ II) \cite{nij-Reid93}.
For the sake of definiteness, however,
we treat ANLV18 as a primary representative.
We shall compare our results with those obtained
with the use of the more {\it traditional} potentials.

\subsection{Monitoring the reliability of the model}

Although, as mentioned, there is by now 
a rather long list of experimental and theoretical work
that points to the basic robustness of 
PhLA calculations, it is desirable to monitor
the reliability of our model by simultaneously studying
reactions that are closely related to the $\nu-d$ reactions
and for which experimental data is available.
It turns out that the $\pi N\Delta$ vertex that features 
in the dominant exchange current for the $\nu-d$ reaction
appears also in the $np\rightarrow \gamma d$ reaction,
for which experimental cross sections are known for 
a wide range of the incident energy, 
from the thermal neutron energy
up to the pion-production threshold.
We therefore calculate here both $\nu-d$ reaction and 
$np\rightarrow \gamma d$ cross sections
in the same formalism and use the latter 
to gauge (at least partially) 
the reliability of our model.

\section{Calculational methods}

\subsection{Multipole expansion of hadron current}

To evaluate the two-nucleon matrix element of 
the hadron current, we first separate the center-of-mass 
and relative wave functions,
\begin{equation}
<\bbox{r}_1,\bbox{r}_2\ |\ d(P)>  
=  e^{i\bbox{P}\cdot\bbox{R}} 
            \psi_d(\bbox{r}),\;\;\;\;\; 
<\bbox{r}_1,\bbox{r}_2\ |\ NN(P')>  
=  e^{i\bbox{P'}\cdot\bbox{R}} 
\psi_{\bbox{p}'}(\bbox{r}),  \label{eq_NN-final}
\end{equation}
where
  $\bbox{r}=\bbox{r}_1 - \bbox{r}_2$ and
  $\bbox{R}=\frac{\bbox{r}_1 + \bbox{r}_2}{2}$,
and $\psi_d$ and $\psi_{\bbox{p}'}$ represent,
respectively, the deuteron wave function
and a scattering-state wave function
with asymptotic relative momentum $\bbox{p}'$.
Then the matrix element of the hadron current for charged-current
  reaction is given by
\begin{eqnarray}
  j^{CC}_{\lambda} \equiv 
 <NN(P)'\ |\ J^{CC}_{\lambda}(0)\ |\ d(P)>
 &=&\int d \bbox{r}\ \psi_{\bbox{p}'}^*(\bbox{r}) 
 \left[ \int d
  \bbox{R}\ e^{-i\bbox{q}  \cdot
  \bbox{R}}   J^{CC}_{\lambda}(0) \right] \psi_d(\bbox{r}).
  \label{eq_j-def}
\end{eqnarray}
As for the neutral-current reaction, 
we just replace $J_\lambda^{CC}$ 
with $J_\lambda^{NC}$.
In the following equations,  
$J_\lambda$ without superscript applies for both NC and CC.
Eliminating the dependence of the current $J_{\lambda}(\bbox{x})$ 
on the center-of-mass coordinate, $\bbox{R}$, we can write
\begin{eqnarray}
  j_{\lambda} = <\psi_{p'}|\!\int d\bbox{x}\ 
   e^{i\bbox{q}\cdot\bbox{x}}
   {\cal J}_{\lambda}(\bbox{x})|\psi_d>,
\end{eqnarray}
where ${\cal J}_{\lambda}(\bbox{x})\equiv 
J_{\lambda}(\bbox{x})|_{\bbox{R}=0}$.
Similarly, we define ${\cal V}_{\lambda}(\bbox{x})\equiv
V_{\lambda}(\bbox{x})|_{\bbox{R}=0}$,
and ${\cal A}_{\lambda}(\bbox{x})
\equiv A_{\lambda}(\bbox{x})|_{\bbox{R}=0}$.
We now introduce the standard multipole expansion 
of the nuclear currents \cite{walecka}.
The multipole operator for the time component of 
a current is defined by
\begin{eqnarray}
T_C^{JM}({\cal J}) & = & \int d\bbox{x} j_J(q
x)Y_{JM}(\hat{\bbox{x}}){\cal J}_0(\bbox{x}),
\label{eq_op-charge}
\end{eqnarray}
where
$j_J(qx)$ is the spherical Bessel function of order $J$,
$q \equiv |\bbox{q}|$, and
$\hat{\bbox{x}}\equiv \bbox{x}/|\bbox{x}|$.
The electric and magnetic multipole operators are
defined by
\begin{eqnarray}
T_E^{JM}({\cal J}) & = & 
  \frac{1}{q} \int d\bbox{x} 
  \bbox{\nabla}\times [j_J(q x)\bbox{Y}_{JJM}
(\hat{\bbox{x}})]    \cdot \bbox{{\cal J}}(\bbox{x}), \\
T_M^{JM}({\cal J}) & = & 
   \int d\bbox{x} 
  j_J(q x)\bbox{Y}_{JJM}(\hat{\bbox{x}})
   \cdot \bbox{{\cal J}}(\bbox{x}),
 \end{eqnarray}
where $\bbox{Y}_{JLM}(\hat{\bbox{x}})$ 
are vector spherical harmonics.
The longitudinal multipole operator is
defined by
\begin{eqnarray}
  T_L^{JM}({\cal J}) & = & 
  \frac{i}{q} \int d\bbox{x} 
  \bbox{\nabla} [j_J(q x)Y_{JM}(\hat{\bbox{x}})]
   \cdot \bbox{{\cal J}}(\bbox{x}). \label{eq_op-longi}
\end{eqnarray}
Using the conservation of the vector current, 
the longitudinal multipole operator of the vector current 
can be related to the charge density operator as
\begin{eqnarray}
T_{L}^{J_o}({\cal V}) = -\frac{\omega}{q}T_{C}^{J_o}({\cal V}).
\end{eqnarray}
An explicit form of the electric multipole operator 
for the vector current is given by
\begin{eqnarray}
  T_{E}^{JM}({\cal V}) &=& -i \sqrt{\frac{J}{2J+1}} \int d
  \bbox{x}j_{J+1}(q x)\bbox{Y}_{JJ+1M}(\bbox{\hat{x}})\cdot
  \bbox{{\cal V}}(\bbox{x}) \nonumber\\
  &&+i \sqrt{\frac{J+1}{2J+1}} \int d
  \bbox{x}j_{J-1}(q x)\bbox{Y}_{JJ-1M}(\bbox{\hat{x}})\cdot
  \bbox{{\cal V}}(\bbox{x}).\label{eq_no-Siegert} 
\end{eqnarray}
Here again we can use the current conservation
to rewrite Eq.(\ref{eq_no-Siegert}) into a form 
that has the correct long wavelength limit
 of an electric multipole operator:
\begin{eqnarray}
  T_{E}^{JM}({\cal V}) &=& 
 -  \sqrt{\frac{J+1}{J}}\frac{\omega}{q}T_{C}^{JM}({\cal V})
-i \sqrt{\frac{2J+1}{J}} \int d
  \bbox{x}j_{J+1}(q x)\bbox{Y}_{JJ+1M}
(\bbox{\hat{x}})\cdot \bbox{{\cal V}}(\bbox{x}).
\label{eq_Siegert} 
\end{eqnarray}

\subsection{Cross sections}

As explained earlier, we calculate
the cross sections
for $\nu/\bar{\nu}(k) + d(P) 
\rightarrow l(k') + N_1(p_1') + N_2(p_2')$
in the laboratory system.
Following the standard procedure, 
we obtain the cross section for the
CC reaction as
\begin{eqnarray}
d\sigma = \sum_{\bar{i},f} 
          \frac{\delta^4(k+P-k'-P')}{(2\pi)^5} 
          \frac{G_F^2 \cos^2\theta_C }{2}\,
           F(Z,E'_{\ell})\,\,|l^ \lambda j_{\lambda}^{CC}|^2 
          d\bbox{k'} d\bbox{p'}_1d\bbox{p'}_2, 
   \label{eq_cs-CC}
\end{eqnarray}
and the cross section for the NC reaction as
\begin{eqnarray}
d\sigma = \sum_{\bar{i},f} 
          \frac{\delta^4(k+P-k'-P')}
     {(2\pi)^5} \frac{G_F^2}{2}|l^
          \lambda j_{\lambda}^{NC}|^2 d\bbox{k'} 
           d\bbox{p'}_1d\bbox{p'}_2.\label{eq_cs-NC}
\end{eqnarray}
The matrix elements, $l^{\lambda}$ and $j_{\lambda}$,
have been defined in Eq.(\ref{eq_l-def}) 
and in Eq.(\ref{eq_j-def}), respectively.
In Eq.(\ref{eq_cs-CC}), we have included
the Fermi function $F(Z,E'_{\ell})$ \cite{fermi-fn} 
to take into account the Coulomb interaction 
between the electron and the nucleons.
In fact, this factor is relevant only
to the $\nu_e+d\rightarrow e^-+p+p$ reaction,
for which we should use $F(Z=2,E'_{\ell})$;
for the $\bar{\nu}_e+d\rightarrow e^++n+n$ reaction
we have $F(Z=0,E'_{\ell})\equiv 1$.

Substitution of the multipole operators 
defined in Eq.(\ref{eq_op-charge})
$\sim$Eq.(\ref{eq_op-longi}) 
leads to
\begin{eqnarray}
  l^ \lambda j_ \lambda &=&
\sum_{J_o M_o} 4 \pi i^{J_o}
  (-1)^{M_o}\nonumber \\
  && \times <\!\psi_{\bbox{p'}}|\left[ 
T_C^{J_oM_o} \ell_C^{J_o-M_o}
  +T_E^{J_oM_o} 
  \ell_E^{J_o-M_o} +
T_L^{J_oM_o} \ell_L^{J_o-M_o}
  +T_M^{J_oM_o} 
\ell_M^{J_o-M_o} \right]\,|\psi_d\!>,
\label{eq_me-jl}
\end{eqnarray}
where the lepton matrix elements are given as
\begin{eqnarray}
  \ell_C^{JM} & = & 
Y_{JM}(\hat{\bbox{q}}) \  l^0,\\
  \ell_E^{JM} & = & 
\left( \sqrt{\frac{J+1}{2J+1}}\bbox{Y}_{J-1 J
  M}(\hat{\bbox{q}})+ 
\sqrt{\frac{J}{2J+1}}\bbox{Y}_{J+1 JM}
(\hat{\bbox{q}}) \right)\cdot \bbox{l},\\
\ell_M^{JM} & = &
\bbox{Y}_{JJM}(\hat{\bbox{q}}) \cdot \bbox{l},\\
  \ell_L^{JM} & = & 
\left( \sqrt{\frac{J}{2J+1}}\bbox{Y}_{J-1 J
  M}(\hat{\bbox{q}})- 
\sqrt{\frac{J+1}{2J+1}}\bbox{Y}_{J+1 J
  M}(\hat{\bbox{q}}) \right)\cdot \bbox{l}.
\end{eqnarray}
To proceed, we use a scattering wave function
of the following form:
\begin{eqnarray}
\psi_{\bbox{p'}}(\bbox{r}) & = & \sum_{L,S,J,T}
 4\pi(1/2,s_1,1/2,s_2|S \mu)\ 
(1/2,\tau_1,1/2,\tau_2|T,T_z)\ 
(L m S \mu |J M) \nonumber\\
&& \;\;\;\;\;\;\;\;\;\;\;\;
\times \,\,i^L Y_{L,m}^*(\hat{\bbox{p}}')\ 
 \psi_{LSJT}(\bbox{r})\label{eq_NNwave}
\end{eqnarray}
with
\begin{eqnarray}
 \psi_{LSJT}(\bbox{r}) & = & 
\frac{1-(-1)^{L+S+T}}{\sqrt{2}}\sum_{L'}
 {\cal Y}_{L'SJ}(\hat{\bbox{r}})
 \ R_{L',L;S}^J (r)\ \eta_{T,T_z},\\
 {\cal Y}_{LSJ}(\hat{\bbox{r}})&=&
\bigl[Y_L(\hat{\bbox{r}})\otimes
 \chi_S \bigr]_{(J)},
\end{eqnarray}
where $\chi_S$ ($\eta_{T}$) is the two-nucleon
spin (isospin) wave function
with total spin $S$ (isospin $T$).
The above wave function is normalized
in such a manner that,
in the plane wave limit, it satisfies 
\begin{eqnarray}
R_{L',L;S}^J(r) \rightarrow j_L(p'r) \delta_{L,L'}.
\end{eqnarray}
The partial wave expansion of the scattering 
wave function [Eq.(\ref{eq_NNwave})] gives
\begin{eqnarray}
  l^ \lambda j_ \lambda   &=& 
\sum_{L,S,J,T,m} \sum_{J_o,M_o}
  (-1)^{M_o} \  i^{J_o-L} 
\frac{(4\pi)^2}{\sqrt{2J+1}}
  (1/2,s_1,1/2,s_2|S \mu)\ 
(1/2,\tau_1,1/2,\tau_2|T,T_z)
  \nonumber\\
  && \times  
(1 m_d J_o M_o  | J M)(L m S \mu |J M) \
  Y_{L,m}(\hat{\bbox{p}}')\sum_{X=C,E,L,M}
< T_X^{J_o} >\ \ell_X^{J_o-M_o},
\end{eqnarray}
where $m_d$ is the z-component of
the deuteron angular momentum.
We have used here a simplified notation
\begin{eqnarray}
<O^{J_o}> & = & <\!\psi_{LSJT} ||  O^{J_o}|| \psi_{d}\! >
\label{eq_simple-rme}
\end{eqnarray}
for the reduced matrix element defined by
\begin{eqnarray}
  <\!J'M'|\ O^{J_oM_o}\ |JM\!> &=& \frac{1}{\sqrt{2J'+1}}\
  (J,M,J_o,M_o|J',M')<\!J'||O^{J_o}||\ J\!>,
\label{eq_rme}
\end{eqnarray}
where $O^{J_oM_o}$ are the multipole operators that appear
in Eq.(\ref{eq_op-charge})$\sim$(\ref{eq_op-longi}).

\subsubsection{Cross sections 
for charged-current reaction}

For the CC reaction, observables of interest are
the total cross section and the lepton 
differential cross sections. 
We therefore integrate Eq.(\ref{eq_cs-CC}) 
over the momenta of the final two nucleons.
The evaluation of the phase space integrals
and the relevant kinematics are briefly described 
in the Appendix.
According to the Appendix, 
Eq.(\ref{eq_cs-CC}) leads to 
\begin{eqnarray}
d\sigma & = &
  \frac{G_F^2 \cos^2 \theta_C}{
  3\pi^2} F(Z,E'_\ell) \,|M|^2
  \delta(M_d+k-E'_{\ell}- P'^0)
  \bar{J}\,p'^2 dp' 
  k'^2 dk' d\Omega_{\bbox{k}'},\label{eq_cs}
\end{eqnarray}
where
\begin{eqnarray}
|M|^2 & = & \sum_{LSJ,J_o} \left\{ \right.
   |<T_{C}^{J_o}({\cal V})>|^2 
\bbox{(}  1+ \hat{\bbox{k}}\cdot\bbox{\beta}
    + \frac{\omega^2}{\bbox{q}^2}\ 
   ( 1- \hat{\bbox{k}}\cdot\bbox{\beta}+2\
   \hat{\bbox{q}}\cdot\bbox{\beta}\  
   \hat{\bbox{q}}\cdot\hat{\bbox{k}})
    - \frac{2 \omega}{q}\ \hat{\bbox{q}}
    \cdot(\hat{\bbox{k}}+\bbox{\beta})\bbox{)}
\nonumber \\
& & + |<T_{C}^{J_o}({\cal A})>|^2 
( 1+ \hat{\bbox{k}}\cdot\bbox{\beta})
    + |<T_{L}^{J_o}({\cal A})>|^2 
( 1- \hat{\bbox{k}}\cdot\bbox{\beta}+2\ 
        \hat{\bbox{q}}\cdot\bbox{\beta}\ 
        \hat{\bbox{q}}\cdot\hat{\bbox{k}}) 
\nonumber \\
& & + 2 Re [<T_{C}^{J_o}({\cal A})><T_{L}^{J_o}({\cal A})>^*]\  
       \hat{\bbox{q}}\cdot(\hat{\bbox{k}}+
\bbox{\beta})  \nonumber \\
& & + [|<T_{M}^{J_o}({\cal V})>|^2 + |<T_{E}^{J_o}({\cal V})>|^2 +
       |<T_{M}^{J_o}({\cal A})>|^2 + |<T_{E}^{J_o}({\cal A})>|^2]\ 
      (1-\hat{\bbox{q}}\cdot\hat{\bbox{k}}\ 
      \hat{\bbox{q}}\cdot\bbox{\beta})  \nonumber \\
& &  \mp 2 Re[<T_{M}^{J_o}({\cal V})><T_{E}^{J_o}({\cal A})>^* + 
              <T_{M}^{J_o}({\cal A})><T_{E}^{J_o}({\cal V})>^*]\ 
      \hat{\bbox{q}}\cdot(\hat{\bbox{k}}-
\bbox{\beta}) \left. \right\}. \label{eq_S-me}
\end{eqnarray}
In the above, $k'\equiv |\bbox{k}'|$ and 
$\bbox{\beta} \equiv \bbox{k}'/E'_{\ell}$;
$\bbox{p}'$ is the relative momentum 
of the final two nucleons, and
$p' \equiv |\bbox{p}'|$.
Of the double sign in the last line of Eq.(\ref{eq_S-me}),
the upper (lower) sign corresponds to 
the $\nu$ ($\bar{\nu}$) reaction.
The appearance of the factor $\bar{J}$
in Eq.(\ref{eq_cs}) needs an explanation.
As discussed in the Appendix,
when relativistic kinematics is adopted,
there arises a Jacobian, $J$, associated with the introduction
of $\bbox{p}'$ but it is a good approximation to use $\bar{J}$,
the angle-averaged value of $J$.

For the total cross section, 
the use of relativistic kinematics gives
\begin{eqnarray}
  \sigma = \int dT \int d(\cos \theta_L) 
  \frac{G_F^2 \cos^2 \theta_C}{3\pi}
 \frac{\bar{J} E'_{\ell} \left(\sqrt{P'^2_{\mu}}/2\right)
 p' k'} {1 + E'_{\ell}(1-k \cos \theta_L/k')/ 
 \sqrt{P'^2_{\mu} + \bbox{q}^2}}
 F(Z,E'_\ell) |M|^2, \label{eq_CC-cs} 
\end{eqnarray}
where $T$ is the kinetic energy of the final NN relative motion
and $\theta_L$ is the lepton scattering angle
($\cos \theta_L=\hat{\bbox{k}}\cdot \hat{\bbox{k}'}$)
in the laboratory frame.
If instead we use non-relativistic kinematics,
the results would be
\begin{eqnarray}
  \sigma = \int dT \int d(\cos \theta_L)
  \frac{G_F^2 \cos^2 \theta_C}{3\pi}
 \frac{E'_{\ell}(2M_r) p' k'}{1 +
 E'_{\ell}(1-k \cos \theta_L/k')/ 
 (M_{N1}+M_{N2})} F(Z,E'_\ell) |M|^2,
\end{eqnarray}
where $M_{Ni}$ is the mass of the $i$-th nucleon,
and $M_r$ is the reduced mass of the final NN system.

Eq.(\ref{eq_cs}) also leads to the double differential 
cross sections for the 
$\nu_e+d\rightarrow e^-+p+p$ reaction:
\begin{eqnarray}
  \frac{d^2 \sigma}{d\Omega_{\bbox{k}'} dE'_{\ell}} &=&  
  \frac{G_F^2 \cos^2 \theta_C}{12\pi^2} 
  F(Z,E'_{\ell})\bar{J}p' k'
  E'_{\ell}\sqrt{P'^2_{\mu} + \bbox{q}^2}\ |M|^2.
  \label{eq_ddx-lep}
\end{eqnarray}
The electron energy spectrum and 
the electron angular distribution are obtained
from Eq.(\ref{eq_ddx-lep}) as
\begin{eqnarray}
 \frac{d \sigma}{dE'_{\ell}}=
  \int d\Omega_{\bbox{k}'}
  \left(\frac{d^2 \sigma}{d\Omega_{\bbox{k}'} 
  dE'_{\ell}}\right)_{\rm Eq.(\ref{eq_ddx-lep})}
\,\,\,\,\,\,\,\,\,\,\,
  \frac{d \sigma}{d\Omega_{\bbox{k}'}} =
  \int dE'_{\ell}\left(\frac{d^2 \sigma}
  {d\Omega_{\bbox{k}'} dE'_{\ell}}\right)_{\rm Eq.(\ref{eq_ddx-lep})}.
  \label{eq_cs-lepspec}
\end{eqnarray}

\subsubsection{Cross sections for neutral-current reaction}

The total cross section for the NC reaction 
can be calculated in essentially the same manner 
as above.  The result is
\begin{eqnarray}
\sigma = \int
d T \int
d(\cos \theta_L) \frac{2 G_F^2}{3\pi}
 \frac{\bar{J}E'_{\ell}\left(
 \sqrt{P'^2_{\mu}}/2\right)
 p' k'}{1 +
 E'_{\ell}(1-k \cos \theta_L/k')/ 
 \sqrt{P'^2_{\mu} + \bbox{q}^2}}
 |M|^2\,, \label{eq_NC-cs} 
\end{eqnarray}
where $|M|^2$ is given by Eq.(\ref{eq_S-me})
with, however, the charged current
replaced by the neutral current. 
By contrast, in calculating neutron 
differential cross sections
we can no longer integrate over
the relative momentum of the final nucleons. 
We therefore work with the following expressions:
\begin{eqnarray}
  \frac{d^2\sigma}{d\Omega_{\bbox{p}'_n} dT_n} = \int
  d\Omega_{\bbox{k}'}\frac{G_F^2}
{3 (2\pi)^5}\frac{E_p
           k'^2p'_nE_n }
           {E_p-\bbox{p}'_p \cdot
           \bbox{\hat{k}}'} 
         \sum_{m_d,s_n,s_p} 
         |j_{\lambda}l^{\lambda}|^2,
        \label{eq_ddx-n}
\end{eqnarray}
where we have indicated explicitly averaging 
over the initial spin 
and summing over the final spins.
The energy and momentum of the final proton (neutron)
are denoted by
$(E'_{\alpha}, \bbox{p'}_{\alpha})$ with $\alpha=p$
($\alpha=n$); 
$T_n$ is the kinetic energy of the neutron.
The neutron energy spectrum and 
the neutron angular distribution are then evaluated as
\begin{eqnarray}
  \frac{d \sigma}{dT_n} =
  \int d\Omega_{\bbox{p}'_n}
  \left(\frac{d^2 \sigma}
  {d\Omega_{\bbox{p}'_n} dT_n}\right)_{\rm Eq.(\ref{eq_ddx-n})}
\,\,\,\,\,\,\,
  \frac{d \sigma}{d\Omega_{\bbox{p}'_n}} =
  \int dT_n \left(\frac{d^2 \sigma}
  {d\Omega_{\bbox{p}'_n} dT_n}\right)_{\rm Eq.(\ref{eq_ddx-n})}.
  \label{eq_cs-nspec}
\end{eqnarray}

The calculation of the total cross section 
for the $np\rightarrow \gamma d$ reaction
follows essentially the same pattern as that
of the $\nu-d$ total cross section,
and therefore we forgo its description.

\section{Numerical results}

\subsection{Radiative capture of neutron on proton}

To test the nuclear currents and wave functions used, 
we first discuss the capture rate for
$np\rightarrow \gamma d$.
Thermal neutron capture is a well known case
for testing exchange currents \cite{riska-brown,fm}.
This reaction is dominated 
by the isovector magnetic dipole transition
from the $^1S_0$ $np$ scattering state. 
With the use of the ANLV18 potential, 
our PhLA calculation gives 
$\sigma(np\rightarrow \gamma d)=335.1$ mb,
with both the IA and MEX currents included.
This is in good agreement with the experimental value,
$\sigma(np\rightarrow \gamma d)^{exp}=
334.2 \pm 0.5$ mb \cite{cox}.
With the IA contribution alone, 
our result would be 
$\sigma(np\rightarrow \gamma d)^{IA}=304.5$ mb.
The 10\% contribution of the exchange current is due to
the pion, pair and $\Delta$-currents. 

\begin{figure}
  \centerline{\epsfig{file=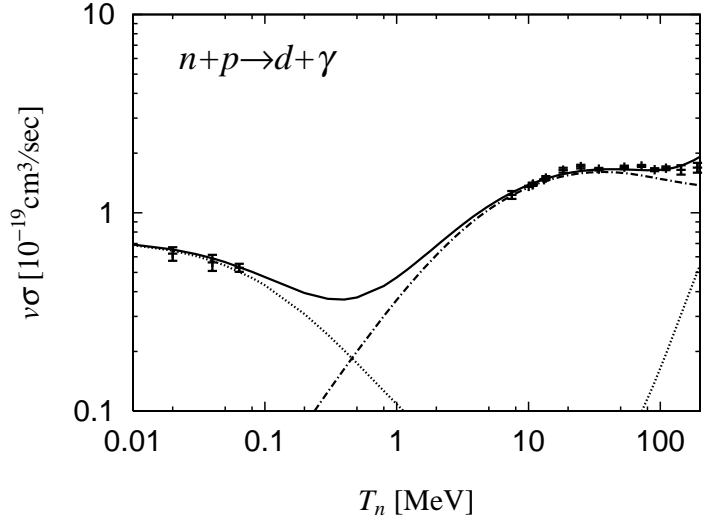,width=10cm}}\vspace{5mm}
  \caption[]{Total cross section for radiative neutron capture.
 The solid curve corresponds to the results of our full calculation
 including the IA and exchange currents and
 all the multipole amplitudes. 
 The dashed and dash-dotted curves show 
 the individual contributions of the magnetic-dipole and 
 electric-dipole amplitudes, respectively.
The data are taken either from the neutron capture reaction
itself \cite{nagai},
or from its inverse process \cite{gamma-d1,gamma-d2},
with the use of detailed balance for the latter.}
    \label{fig_neu-cap}
\end{figure}

Going beyond the thermal neutron energy regime,
we give in Fig.\ref{fig_neu-cap} the calculated 
$\sigma(np\rightarrow \gamma d)$
as a function of the incident neutron kinetic energy, 
$T_n$.
The experimental data in Fig.\ref{fig_neu-cap} 
have been obtained from either 
the neutron capture reaction itself \cite{nagai}
or its inverse process \cite{gamma-d1,gamma-d2},
using detailed balance for the latter. 
We can see that our results describe very well
the energy dependence of 
$\sigma(np\rightarrow \gamma d)^{exp}$
all the way up to $T_n\approx$ 100 MeV.
The figure indicates that
the electric dipole amplitude starts to become important
around $T_n$=100 keV.
In the higher energy region we should expect deviations from
the long-wave length limit of the electric dipole operator,
and therefore the good agreement of our results 
with the data suggests that
the description of the electric multipole is 
also satisfactory.\footnote{Since our treatment 
here does not include pion production, 
our results should be taken with caution
above the pion production threshold.
}
The fact that our PhLA calculation
with no {\it ad hoc} adjustment of the input parameters
is capable of reproducing 
$\sigma(np\rightarrow \gamma d)^{exp}$
for a very wide range of the incident energy
gives us a reasonable degree of confidence 
in the basic idea of PhLA
and the input parameters used.\footnote{
Another similar success of PhLA is known
in the $d(e,e')np$ reaction \cite{fm}.}
Of course, strictly speaking, 
the electromagnetic and weak-interaction processes
do not probe exactly the same sectors of PhLA,
but the remarkable success with 
$\sigma(np\rightarrow \gamma d)$
gives, at least, partial justification of our PhLA
as applied to weak-interaction reactions.
Noting that the dominant
axial MEX current due to $\Delta$-excitation
is related to the $\Delta$-excitation MEX current 
for the vector current 
(we need only replace $(f_V+f_M)/2M_N$ with $f_A$ ),
we evaluate the former
with the same input parameters
as used in calculating $\sigma(np\rightarrow \gamma d)$.

\subsection{Cross sections of $\nu-d$ reactions}

We now present our numerical results 
for the $\nu(\bar{\nu})-d$ reactions.
In what follows, the {\it ``standard run"}
represents our full calculation 
with the following features.
The ANLV18 potential \cite{anlv18} is used to generate
the initial and final two-nucleon states
and the final two-nucleon partial waves
are included up to $J=6$.  
For the transition operators, we use the IA and 
MEX operators described in Sec. \ref{section_form};
the Siegert theorem is invoked for the 
electric part of the vector current.
As regards the single-nucleon weak-interaction form factors,
we employ the most updated parameterization given
in Eqs.(\ref{eq:fv})-(\ref{eq:GA}).
The final two-nucleon system is treated relativistically
in the sense explained in Appendix.\footnote{
We must emphasize that our calculation 
takes account of ``relativity" only in certain aspects 
of kinematics.  Going beyond this is out of the scope of this paper.}
Our numerical results will be given primarily 
for our {\it standard run}; other cases are 
presented mostly in the context of examining 
the model dependence.

\subsubsection{Total cross sections for $\nu-d$
and $\bar{\nu}-d$ reactions}

\begin{figure}
  \centerline{\epsfig{file=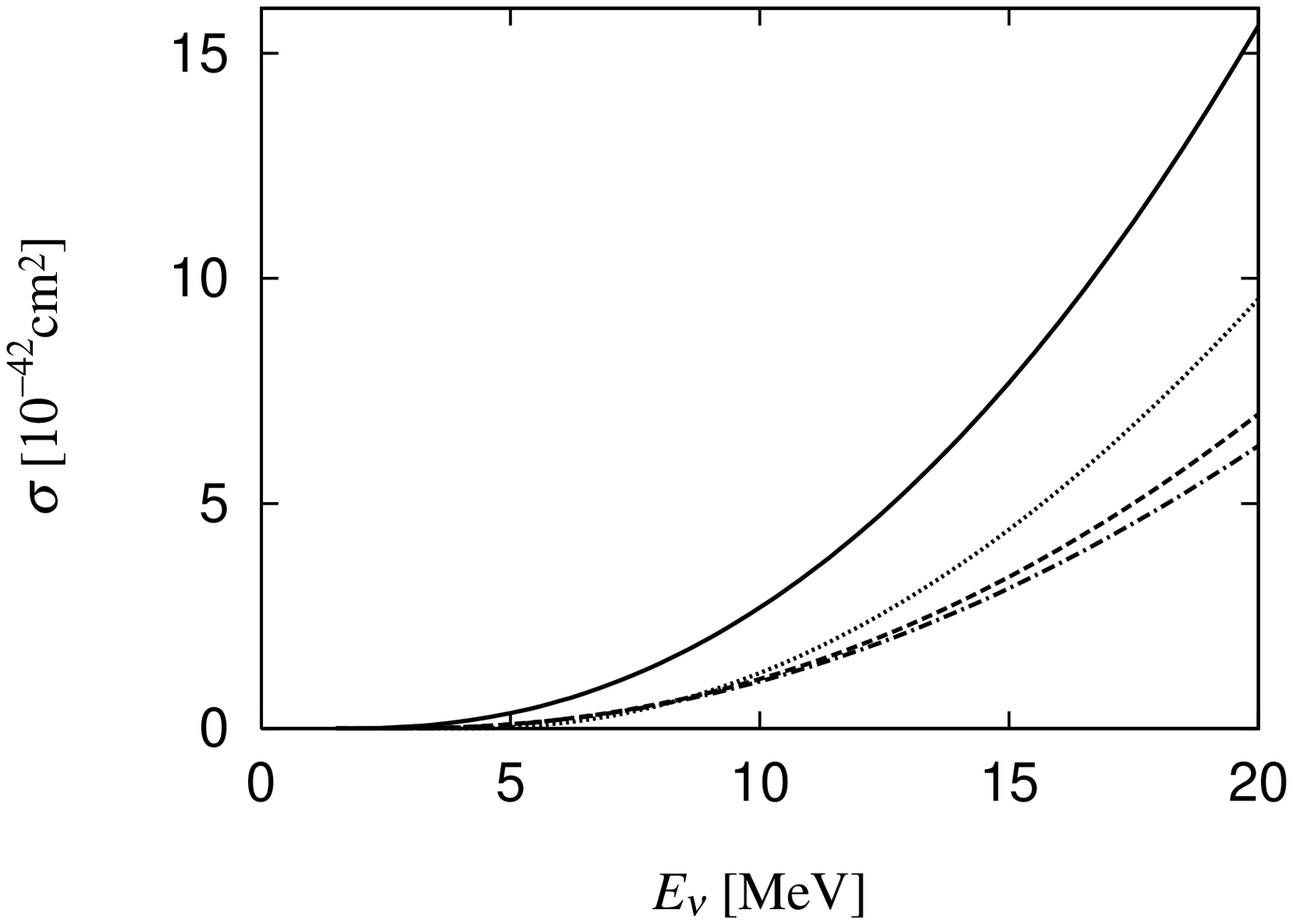,width=10cm}}
  \centerline{\epsfig{file=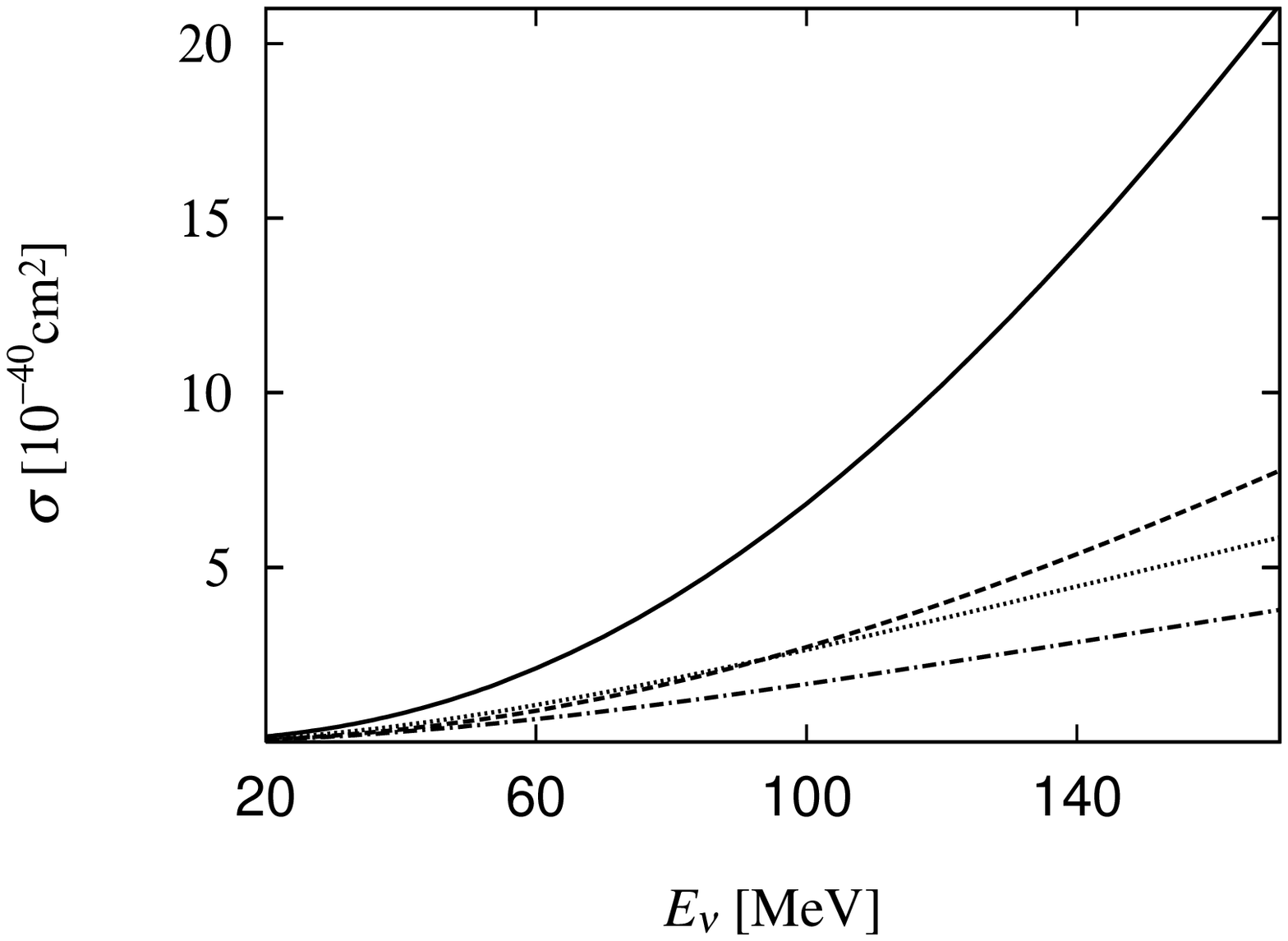,width=10cm}}
  \vspace{5mm}
  \caption[]{Total cross section for the reactions:
  $\nu_e d \rightarrow e^-pp$,\,
  $\bar{\nu}_e d \rightarrow e^+nn$,\, 
  $\nu d \rightarrow \nu pn$, and
  $\bar{\nu} d \rightarrow \bar{\nu}pn$.
  The solid and dotted curves show 
 the charged-current reaction cross sections 
 for $\nu$ and $\bar{\nu}$,  
 respectively, while 
 the long-dash and dash-dotted curves give 
 the neutral-current reaction cross sections 
 for $\nu$ and $\bar{\nu}$, respectively.}
    \label{fig_tot}
\end{figure}

We give in TABLE \ref{table_tot} and Fig.\ref{fig_tot}
the total cross sections, obtained in our {\it standard run},
for the four reactions:
$\nu_e d\rightarrow e^- pp$,\,
$\nu_x d\rightarrow\nu_x np$,\,
$\bar{\nu}_e d\rightarrow e^+ nn$,\,
$\bar{\nu}_x d\rightarrow\bar{\nu}_x np$.
The cross sections are given as functions of $E_{\nu}$,
the incident $\nu/\bar{\nu}$ energy, 
from the threshold to $E_{\nu}=170$ MeV.
\footnote{The numerical results reported 
in this article are available 
in tabular and graphical forms at the web site:
$<$http://nuc003.psc.sc.edu/ $\tilde{ }\ $kubodera/NU-D-NSGK$>$.
\label{foot_web}}
It should be mentioned that 
towards the highest end of $E_{\nu}$ considered here,
pion production sets in but the present calculation
does not include it.

It is informative to decompose the total cross section
into partial-wave contributions.
TABLE \ref{table_cnf} shows the relative importance of
the two lowest partial waves 
in the final two-nucleon state;
denoting the contributions to the total cross section
from the $^1S_0$ and $^3P_J$ states
by $\sigma(^1S_0)$ and $\sum_J\sigma(^3P_J)$, respectively,
we give in TABLE \ref{table_cnf} the ratios,
$\sigma(^1S_0)/\sigma({\rm all})$ and
$\sum_{J=0}^2\sigma(^3P_J)/\sigma({\rm all})$,
as functions of $E_\nu$.
Here $\sigma({\rm all})$ denotes the sum of the contributions 
of all the partial waves; in fact, it is 
sufficient to include up to $J=6$ 
even for $E_{\nu}=170$ MeV, where
the summed contribution of higher partial waves ($J > 6$) 
is found to be less than 1\%.
The table reconfirms that, in the low-energy region,
the Gamow-Teller (GT) amplitude 
due to the $^1S_0$ final state gives
a dominant contribution.
It is therefore important to take into account 
the $\Delta$-excitation axial-vector current,
which gives a main correction to the IA current.
As mentioned, in our approach,
the coupling constant determining
the $\Delta$-excitation MEX current is 
controlled by the $np\rightarrow \gamma d$ amplitude.
As $E_\nu$ increases, 
the $^3P_J$ final states become
as important as the $^1S_0$ state,
and therefore $1^-$ type multipole operators 
arising from the vector as well as axial-vector currents
start to play a significant role. 
In this sense it is reassuring that the validity 
of our model for the electric dipole matrix element 
in this energy region has been tested in the photo-reaction.

Turning now to TABLE \ref{table_crnt}, we give in the
second column labeled ``IA",
the ratio of the total cross section obtained 
with the use of the IA terms alone to that of 
our {\it standard run}.
We see that, at the low energies, the MEX contribution 
is about 5\% of the IA contribution.
As $E_\nu$ increases, 
the relative importance of the MEX current contribution 
augments and it can reach as much as 8\% 
in the high energy region.
The third column ($+\bbox{A}_{MEX}$) in Table \ref{table_crnt}
gives the cross section
that includes the contribution of 
the space component of the axial exchange current,
while the fourth column ($+A_{KDR,0}$) 
gives the results that contain the additional
contribution of the time component of 
the axial exchange current.
It is clear that the MEX effects are
dominated by $+\bbox{A}_{MEX}$;
the axial-charge contribution is very small
for the entire energy range considered here.
The last column ($+\bbox{V}'_{MEX}$) 
in Table \ref{table_crnt} gives results 
obtained with the use of the full vector exchange currents,
Eq.(\ref{eq_no-Siegert}), 
i.e., without invoking the Siegert theorem.
The numerical difference between the two cases 
(with or without the Siegert theorem imposed) 
is found to be very small;
the difference is practically zero 
for lower values of $E_\nu$
and, even at the higher end of $E_\nu$,
it is less than 1\%.
Thus the Siegert theorem allows us to take into account 
implicitly most part of the MEX 
for the vector current.\footnote{
In our approach, which uses phenomenological nuclear
potentials, the conservation of the vector current
is not strictly satisfied.  A measure of the
effect of current non-conservation may be provided
by comparing two calculations, one with the Siegert 
theorem implemented and the other without.
The results in Table \ref{table_crnt} indicate
that numerical consequences of the current non-conservation
are practically negligible in our case.}

In order to compare our cross sections
with those of the previous work,
we give in TABLE \ref{table_model} the
ratios of the cross sections reported 
in YHH \cite{YHH2} and in KN \cite{kubo1} to
those of our {\it standard run};
the second column gives 
$\sigma({\rm Y\!H\!H})/\sigma(standard\,\,run)$, while 
the third column shows
$\sigma({\rm K\!N})/\sigma(standard\,\,run)$.
In the solar neutrino energy region, one can see that the results of
our {\it standard run} agree with those of KN \cite{kubo1}
within 1\% except for the $\nu_e d \rightarrow e^-pp$ reaction near
threshold, wherein the discrepancy can reach 2\%.  
As the incident energy becomes higher,
our results start to be somewhat larger than 
those of KN, and the difference becomes about 6\% 
towards the higher end of $E_\nu$. 
This variance arises largely from the cutoff mass 
in the form factor $G_A(q^2)$, which accounts for 3-4\% 
difference.\footnote{The value of the cutoff mass, $m_A$, in \cite{avm} 
was deduced from an experiment involving a deuteron 
target and therefore it may involve nuclear effects. 
It seems worthwhile to reanalyze the data
taking into account possible nuclear effects. 
Another potentially useful source of information 
on $m_A$ is low-energy pion electroproduction \cite{MAINZ}.}
The remaining $\sim$2\% difference is due to our use of relativistic
kinematics and the inclusion of the contributions from higher partial
waves and from the isoscalar current which were ignored in the previous study.
We have done an additional calculation
by running our code adopting the same approximations and 
the same input parameters as in KN, 
and confirmed that the results agree with those of KN
within 1\% in the high energy region 
as well.\footnote{The precision of our numerical
computation of the cross sections is also 1\%.}

On the other hand, the cross sections of YHH \cite{YHH2}
are about 5\% smaller than those of our {\it standard run}
even at the low energy.
This reflects the fact that
YHH did not include the MEX contributions
(except for the term that could be incorporated via 
the extended Siegert theorem).
Indeed, comparison of the YHH cross sections
with the entries in the second column labeled ``IA"
in TABLE \ref{table_crnt} indicates that, 
if we drop the explicit MEX terms in our calculation,
the resulting cross sections in the solar energy region
agree with those of YHH within $\sim$1\%.

We next consider the NN-potential dependence 
of the cross sections.
The fourth column labeled ``Reid93"
in TABLE \ref{table_model}
gives the ratio of the total cross section
obtained with the use of the Reid93 potential
\cite{nij-Reid93} to that of our {\it standard run};
the fifth column gives a similar ratio 
for the case of the NIJ II potential \cite{nij-Reid93}.
We note that the dependence on the nuclear potentials 
is within 1\% for all the reactions
and for the entire energy region under study.
\footnote{There is 2\% variance 
for the $\bar{\nu}_e d \rightarrow e^+ nn$ cross section
near threshold (not shown here); 
this is however very likely to be attributable
to the fact that the $n$-$n$ scattering length
is not exactly reproduced by the potentials other than ANLV18.} 
Since all the potentials used here
describe the NN scattering data to a satisfactory degree,
it is probably not extremely surprising  
that all these {\it modern} realistic NN potentials
give essentially identical results for
$\nu-d$ cross sections,
but the present explicit confirmation is reassuring. 

In our calculation the strength of the
$\Delta$-excitation exchange current, 
which contributes both to the
Gamow-Teller and M1 transitions, is monitored by
the empirical values for 
$\sigma(np\rightarrow \gamma d)$.
Meanwhile, Carlson {\it et al.} \cite{crsw},
in estimating the solar $pp$-fusion cross section,
used the tritium $\beta$-decay rate to fine-tune
the $\pi N\Delta$ coupling constant that features
in the Gamow-Teller exchange current.
This method turns out to yield somewhat ``quenched" 
$\Delta$-excitation MEX effects in the $pp$ fusion.
It is therefore of interest to study
the consequences of this second method
for the $\nu-d$ reactions.
In the last column labeled ``$\Delta$(CRSW)"
of TABLE \ref{table_model},
we give the ratio of the cross sections obtained
with the use of the $\Delta$-current employed in
\cite{crsw} to those of our {\it standard run}.
In the solar energy region this ratio is found to be
0.96 - 0.97, or the MEX contribution relative 
to the IA term is 2\%,
instead of 5\% found in our {\it standard run}. 
This reduction is primarily due to the smaller 
$\pi N\Delta$ coupling constant in \cite{crsw}.
At higher neutrino energies, 
the use of the the $\Delta$-current 
employed in \cite{crsw} leads to a $\sim$4\% 
MEX effect relative to the IA term, 
to be compared with the $\sim$8\% effect
found in our {\it standard run}.
Thus, in general, 
if we adopt the approach taken in \cite{crsw},
the importance of the MEX effect relative to 
the IA contribution will be reduced by a factor 
of $\sim$2 as compared with the result of
our {\it standard run}.

As emphasized by Bahcall {\it et al.} \cite{bks00},
one of the crucial quantities
in neutrino oscillation studies at the SNO
is the double ratio $[\mbox{NC}]/[\mbox{CC}]$,
where $[\mbox{NC}]$ ($[\mbox{CC}]$) itself 
is the ratio of the observed neutrino absorption rate 
to the standard theoretical estimate 
for the NC (CC) reaction rate.  
This implies that the reliability of 
theoretical estimates for the ratio
$R \equiv \sigma(NC)/\sigma(CC)
\equiv \sigma(\nu d \rightarrow \nu np)/
\sigma(\nu_e d\rightarrow e^-pp)$
is extremely important.
We give in Table \ref{table_ratio_model}
the values of $R$ resulting from the various models
considered in this paper.
Since our primary interest here is to examine
the model dependence of $R$,
we choose, in Table \ref{table_ratio_model},
to normalize $R$ by $R_{standard\; run}$,
the value corresponding to our {\it standard run};  
$R_{standard\; run}$ itself is shown in the second
column of the table.
We learn from Table \ref{table_ratio_model}
that all the models studied give essentially the same $R$; 
deviations from $R_{standard\; run}$
are at most $\sim$1\%.
Thus, the largest source of model dependence
in our work due to the
$\Delta$-exchange current cancels out 
by taking the ratio between the NC and CC reactions.

\subsubsection{Differential cross sections
for the electron}

\begin{figure}
\centerline{
\epsfig{file=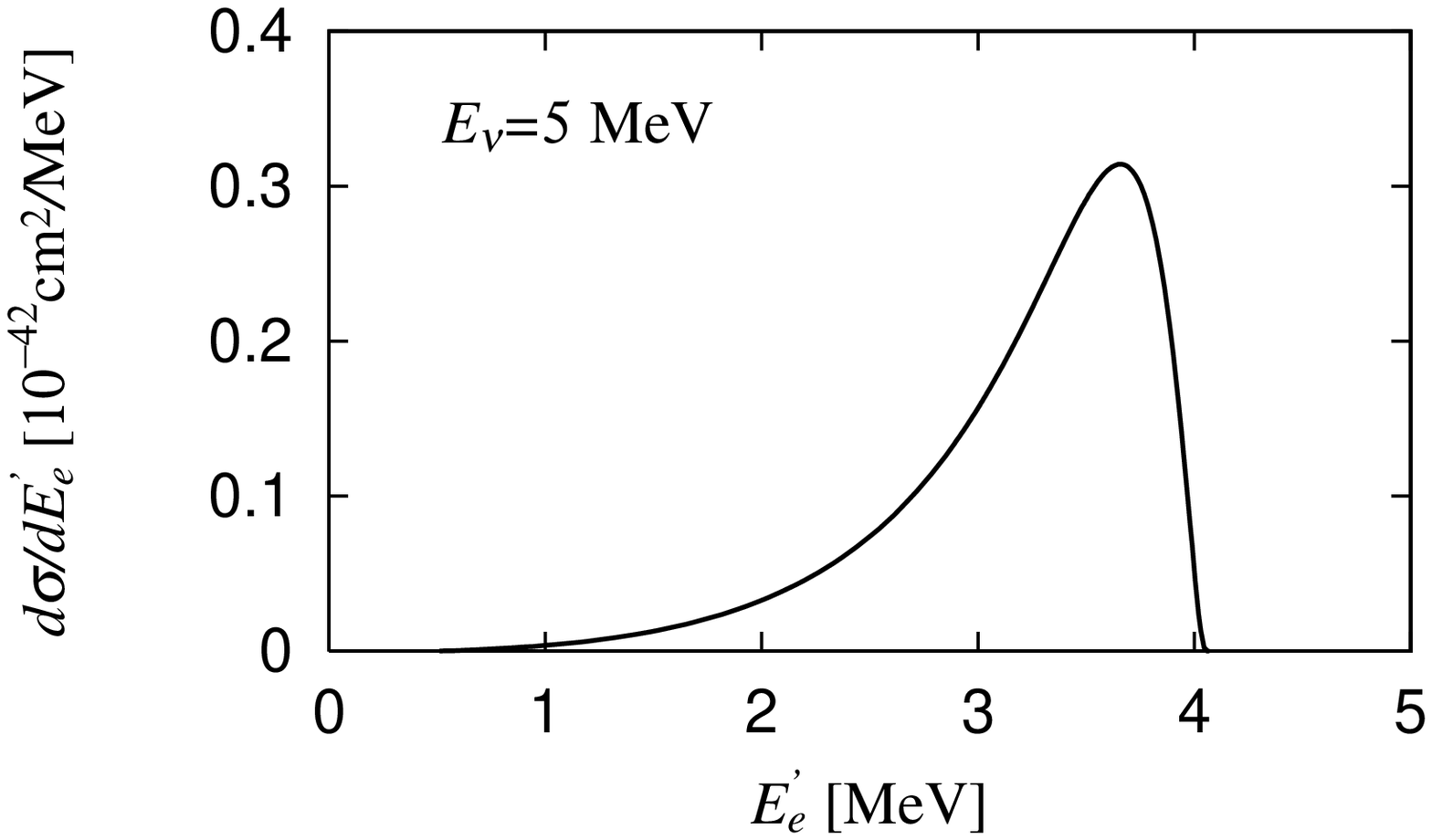,width=9cm}  
}                                      
\centerline{                           
\epsfig{file=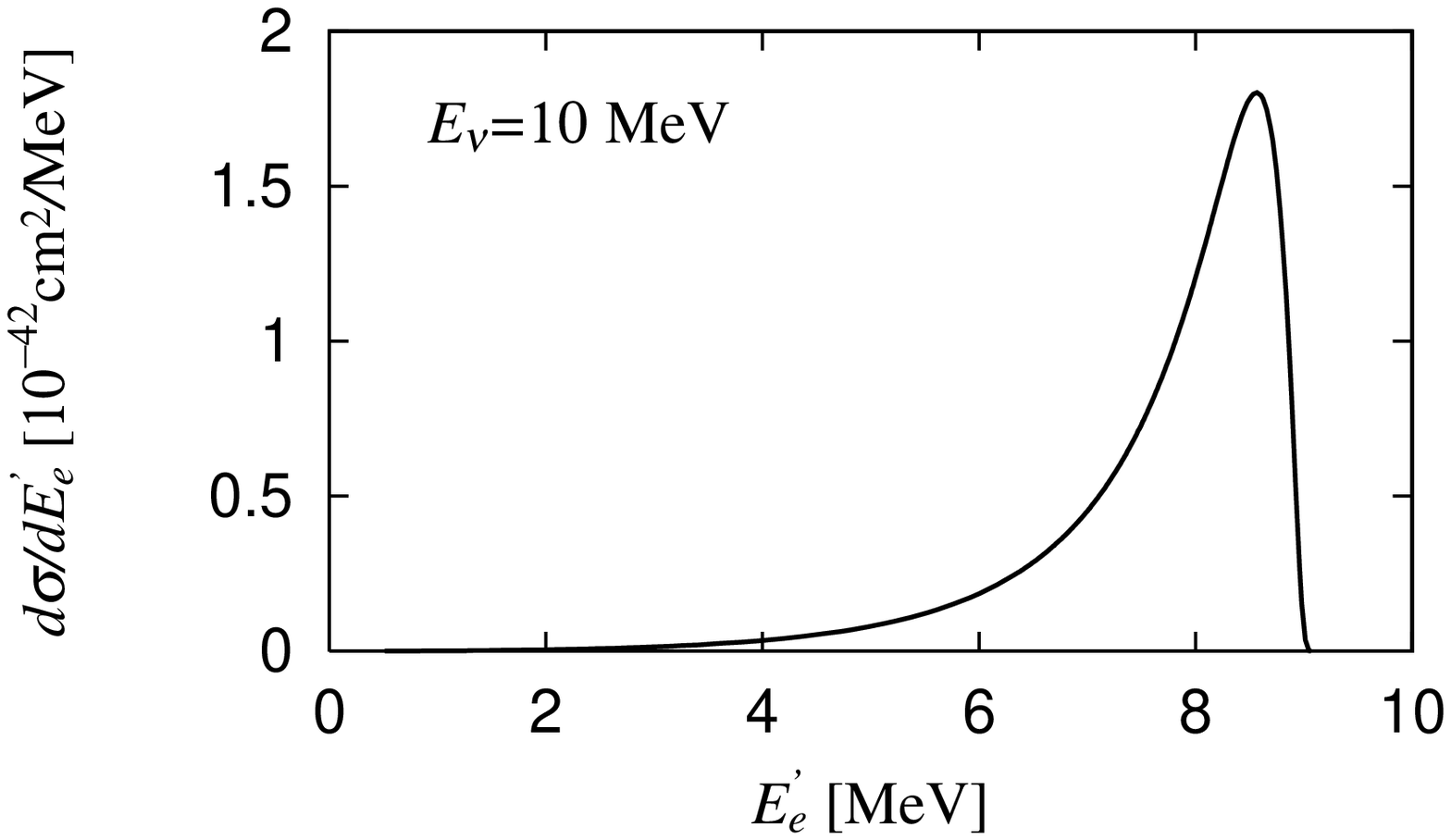,width=9cm}
}                                      
\centerline{                           
\epsfig{file=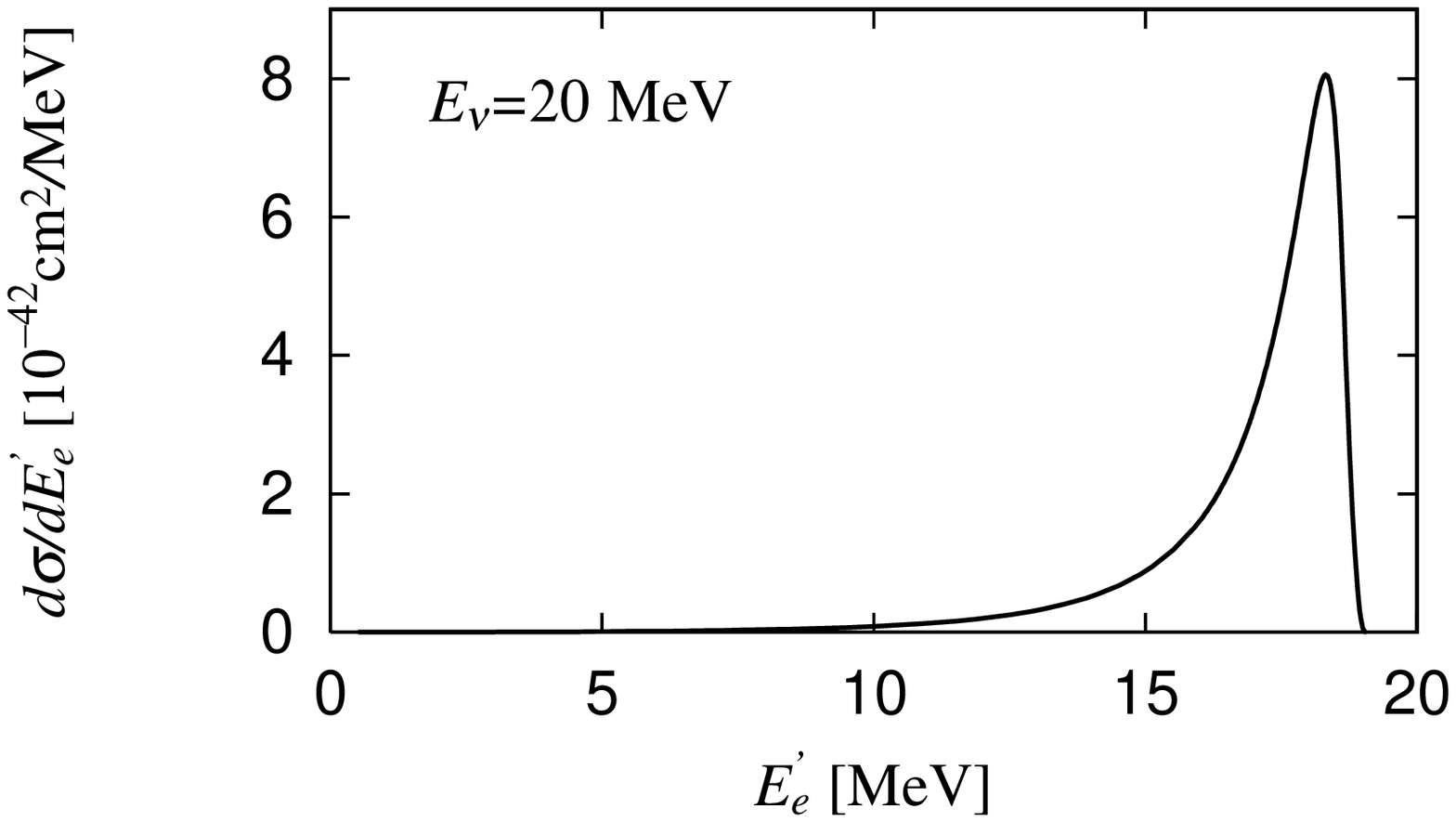,width=9cm}
}                                      
\centerline{                           
\epsfig{file=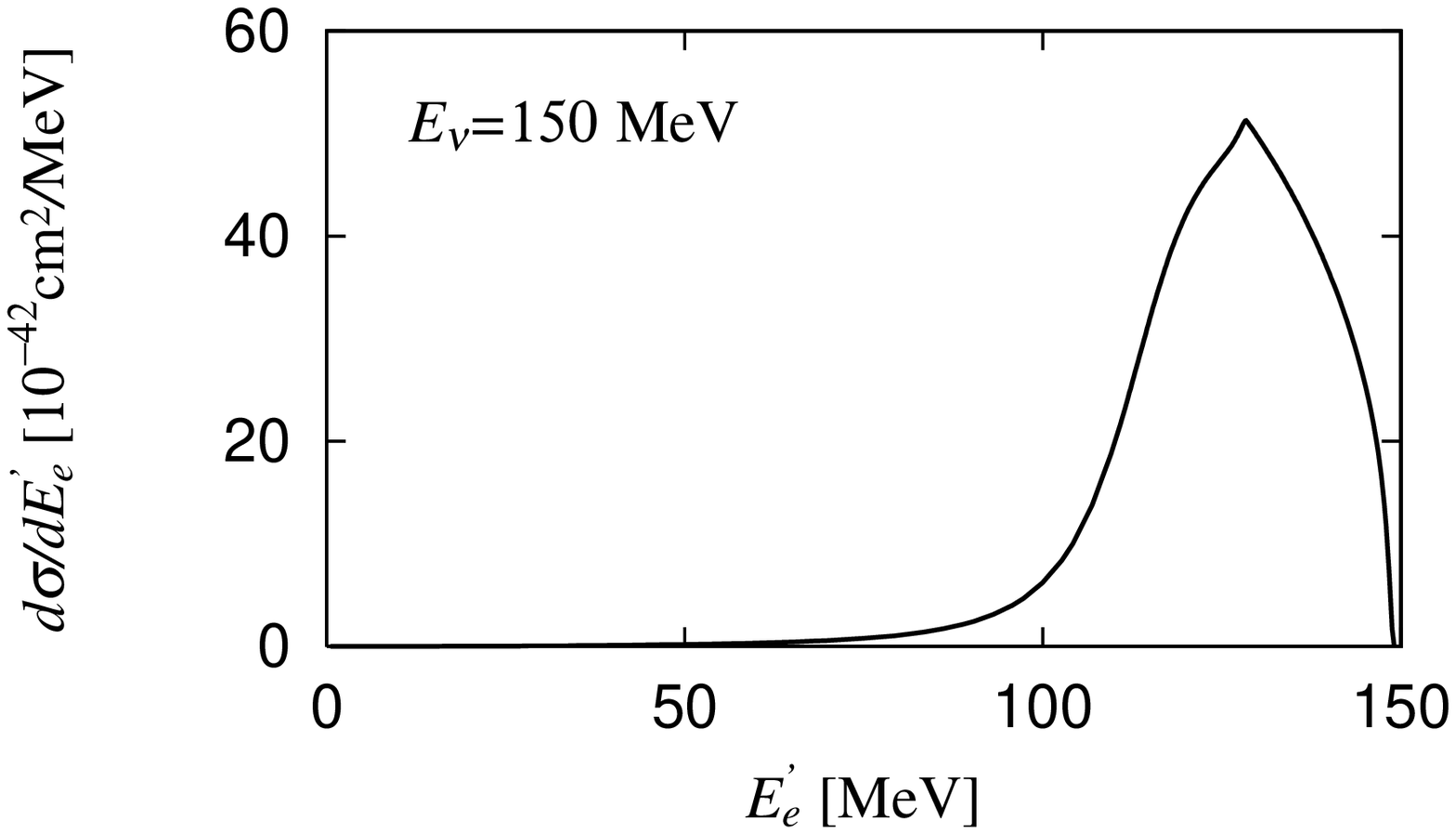,width=9cm}
}  
\caption[]{Electron energy spectra for the 
$\nu_e d \rightarrow e^- pp$ reaction.}
\label{fig_lepspec}
\end{figure}

\begin{figure}

\centerline{
\epsfig{file=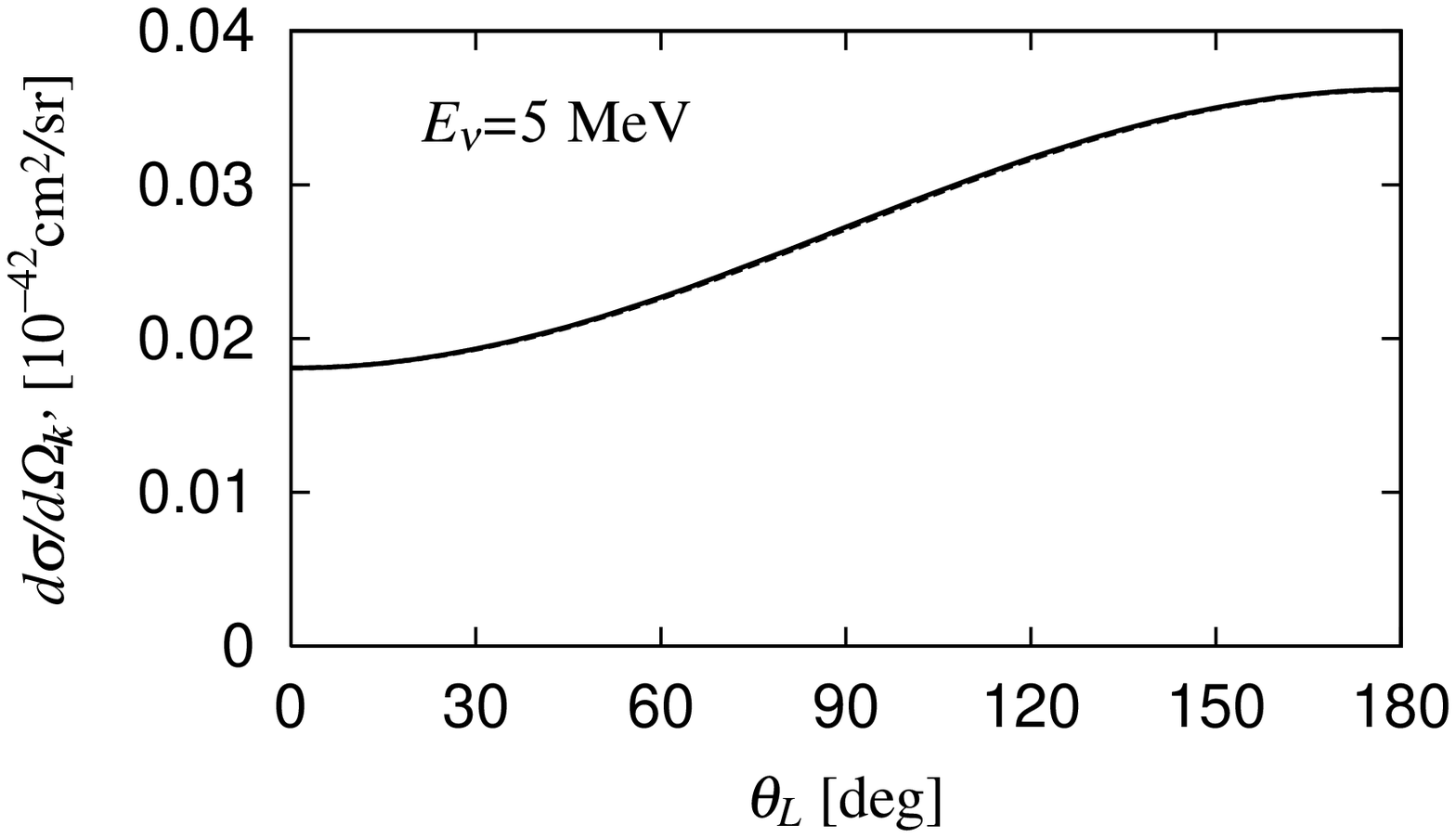,width=9cm}  
}                                      
\centerline{                           
\epsfig{file=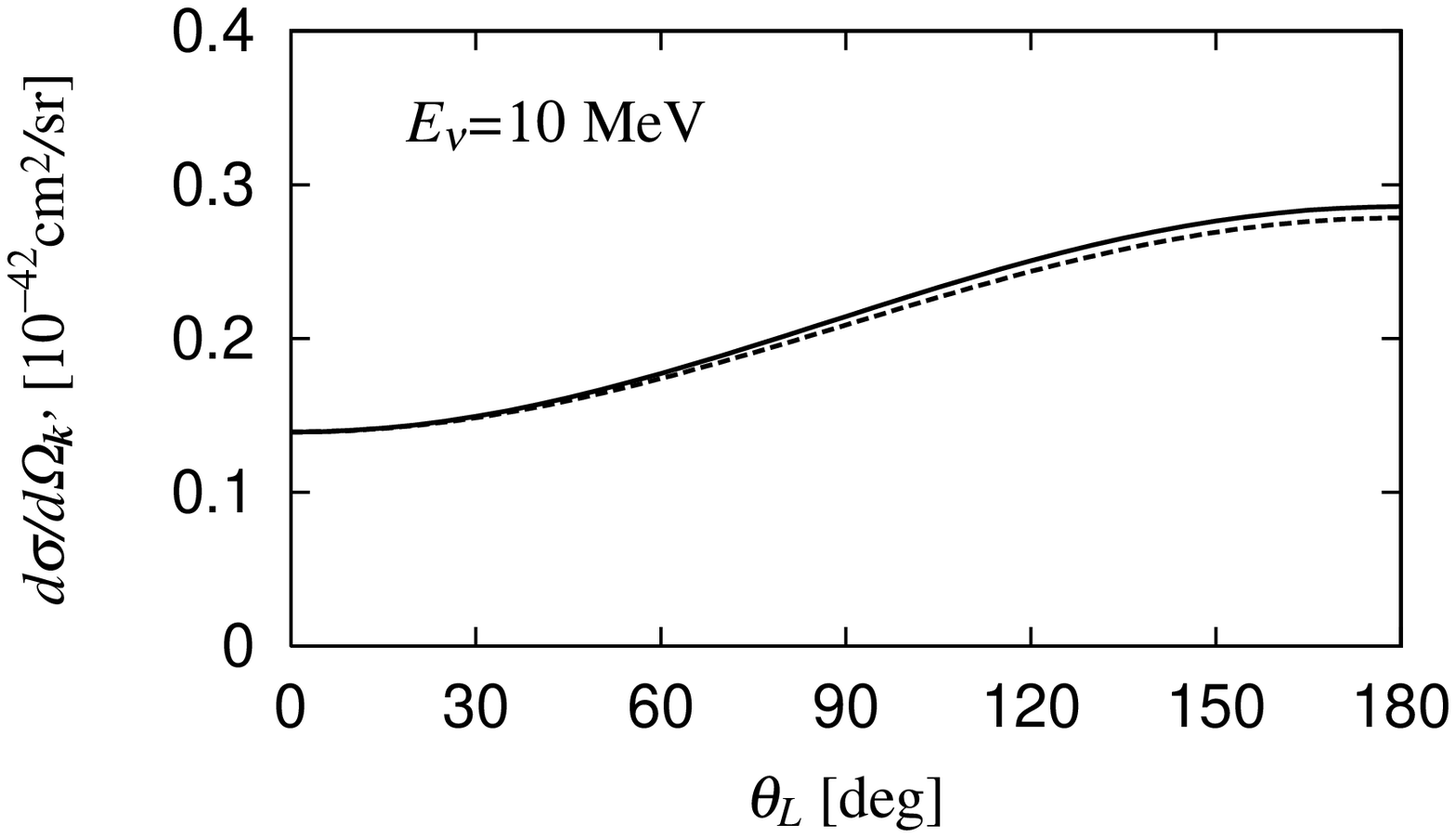,width=9cm} 
}                                      
\centerline{                           
\epsfig{file=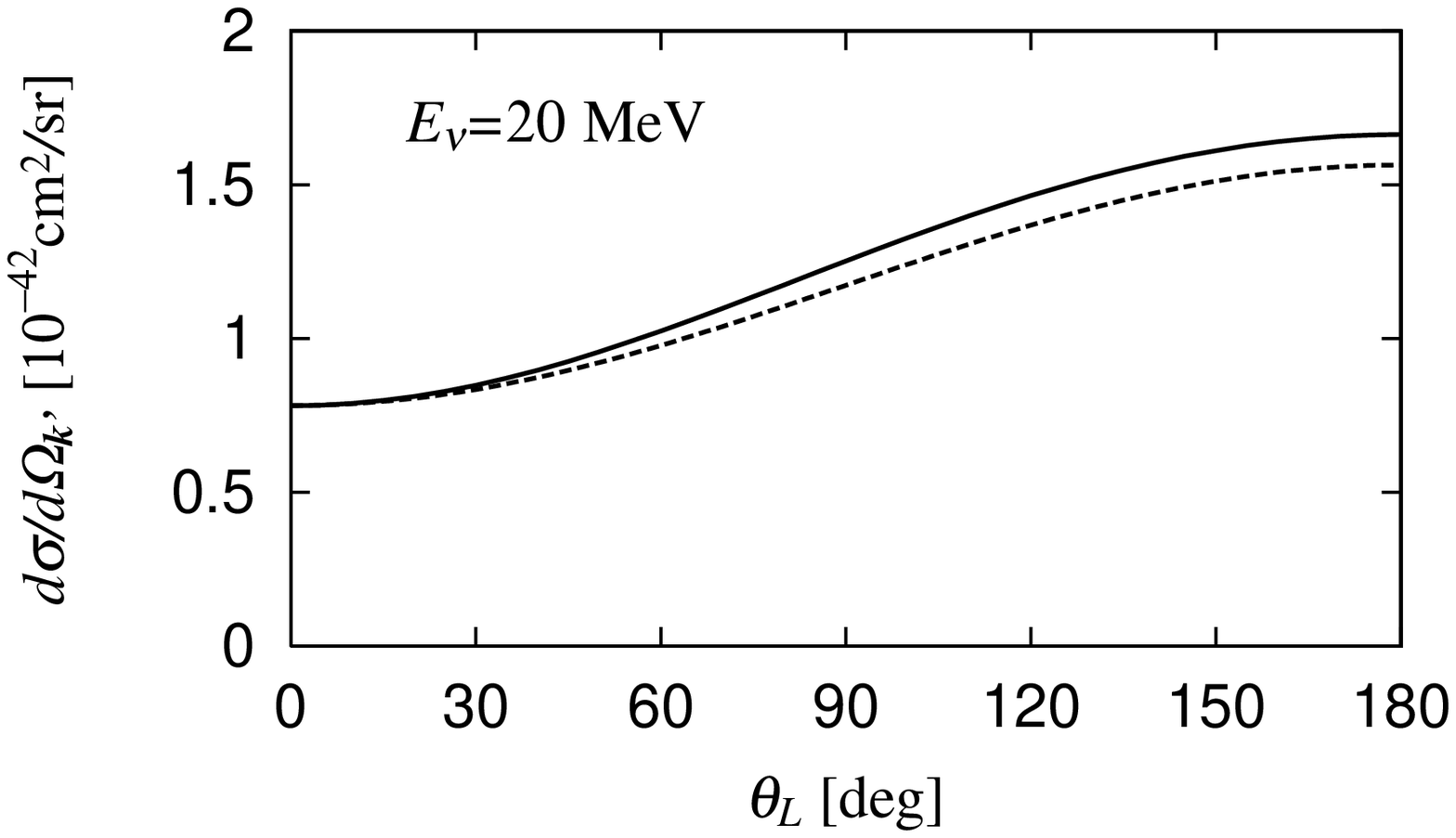,width=9cm} 
}                                      
\centerline{                           
\epsfig{file=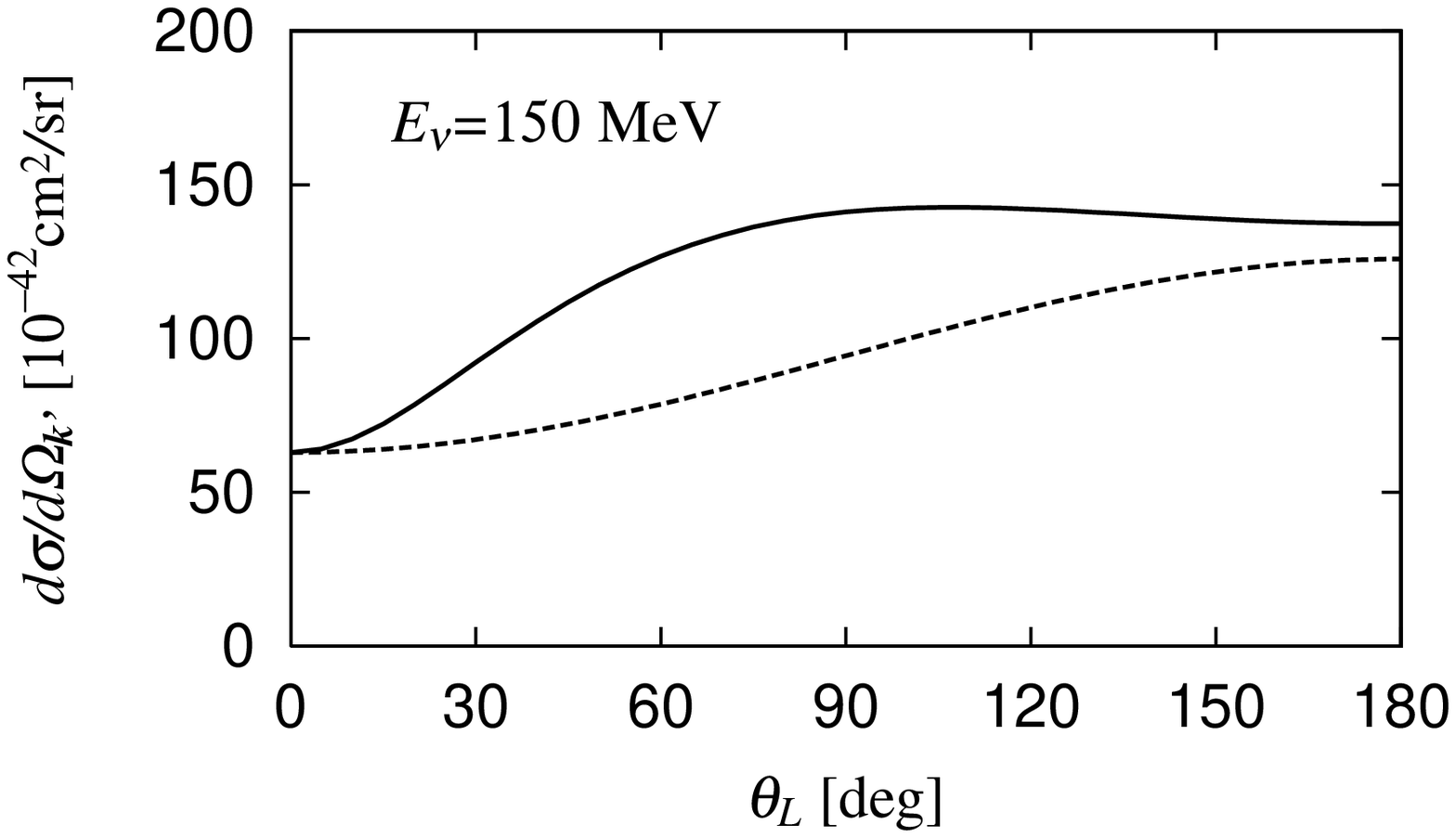,width=9cm}
}
\caption[]{Electron angular distribution 
for the $\nu_e d \rightarrow e^- pp$ reaction. 
The solid curves show the results of our {\it standard run},
while the dotted curves correspond to the simplified expression, 
Eq.(\ref{eq_reduced-lep}),
normalized to the {\it standard run} results
at $\theta_L=0$.}
\label{fig_lepagl}
\end{figure}

We now discuss three types of
electron differential cross sections
for the $\nu_e + d \rightarrow e^- + p + p$ reaction:
(i) the energy spectrum, $d\sigma/dE'_e$
in Eq.(\ref{eq_cs-lepspec}),
(ii) the electron angular distribution, 
$d\sigma/d\Omega_{\bbox{k}'}$ 
in Eq.(\ref{eq_cs-lepspec}),
and (iii) the electron double differential cross sections,
$d^2\sigma/dE'_e d\Omega_{\bbox{k}'}$ in Eq.(\ref{eq_ddx-lep}).
Although this kind of information must be implicitly contained
in the computer codes used in the existing work 
\cite{TKK,DK1,YHH2,kk92,kubo1},
its explicit tabulation has been lacking in the literature.
It seems very useful to make these differential cross
sections readily available to our research community.
However, a trivial but nonetheless serious problem 
is that the required amount of tabulation is enormous.
We therefore present here some representative results,
relegating the bulk of tabulation 
to a web site.\footnote{See footnote \ref{foot_web}.}
For four values of the incident neutrino energies, 
$E_{\nu}$ = 5, 10, 20 and 150 MeV,
we give the electron-energy spectra,
$d\sigma/dE'_e$, 
in Fig.\ref{fig_lepspec}
and the electron angular distribution,
$d\sigma/d\Omega_{\bbox{k}'}$,
in Fig.\ref{fig_lepagl}.
We note that the electron spectrum in Fig.\ref{fig_lepspec} 
exhibits a ``cusp-like" structure for $E_{\nu}$=150 MeV. 
This feature, which is in fact common for $E_{\nu}\geq$ 100 MeV,
probably calls for an explanation.
For a given value of $E_\nu$,
we can separate the electron energy $E'_e$
into two ranges: $E'_e<E'^c_e$ or $E'_e>E'^c_e$,
where $E'^c_e$ is the point above which 
the electron scattering angle $\theta_L$
cannot any longer cover the full range $[0,\pi]$ 
for a kinematic reason.\footnote{See the Appendix.}
The ``cusp-like" structure occurs at $E'_e=E'^c_e$ 
due to interplay between the change 
in the range in the phase space integral and
the momentum dependence in the transition matrix element 
for the final ${}^1S_0$ channel.
This structure, however, is {\it not} a cusp 
in the mathematical sense.  Enlarging the scale of
the abscissa, we can confirm that the actual curve is
a rapidly changing but non-singular one. 
It turns out that for higher values of $E_\nu$
we need more scale enlargement before the curve starts
looking smooth to the eye.  This is the reason why, 
for a fixed abscissa scale 
(as adopted in our illustration),
the case corresponding to the high incident energy
tends to exhibit more ``cusp-like" behavior.

Regarding the electron angular distribution (Fig.\ref{fig_lepagl}),
we note that at low neutrino energies 
the electrons are emitted in the backward direction,
carrying most of the available energy. 
The angular distributions for the lower incident
energies are reminiscent of that for
a Gamow-Teller $\beta$-decay between two bound states.
If we simplify the expression
for the electron differential cross section,
(Eq.(\ref{eq_ddx-lep})), 
by dropping all the partial waves 
other than $^1S_0$ and by retaining only
the leading-order Gamow-Teller matrix element, 
then we have 
\begin{eqnarray}
    d \sigma &\sim&  \frac{G_F^2 \cos^2 
    \theta_C}{12\pi^3}f_A^2M_p
  p' k'^2 F(Z,E'_e)\ 
  ( 3 - \beta \cos \theta_L)
  I^2 dk'd\Omega_{\bbox{k'}},
\end{eqnarray}
where $I$ is the relevant radial integral. 
Since $\beta\sim 1$ and $F\sim 1$,
if we tentatively treat $I$ as a constant,
we have a simple expression
\begin{eqnarray}
  d\sigma &\propto& p'k'^2\ 
  ( 3 -  \cos
  \theta_L)dk'd\Omega_{\bbox{k'}}. 
  \label{eq_reduced-lep}
\end{eqnarray}
In fact, the electron angular distributions 
for low incident neutrino energies
can be simulated to high accuracy 
by Eq.(\ref{eq_reduced-lep});
see the dotted lines in Fig.\ref{fig_lepagl}. 
Thus, although the radial integral $I$ 
may in fact depend strongly on the kinetic energy
of the NN relative motion, the numerical results 
for $d\sigma/d\Omega_{\bbox{k}'}$ at the low energies
can be conveniently simulated by the simple 
phase-space formula, Eq.(\ref{eq_reduced-lep}).

As for the electron double differential cross sections,
$d^2\sigma/dE'_e d\Omega_{\bbox{k}'}$, Eq.(\ref{eq_ddx-lep}),
even presenting some typical cases
is impractical because of the bulkiness of the tables.
We therefore relegate their tabulation completely
to the web site the address of which is given in footnote \ref{foot_web}.

\subsubsection{Neutron energy spectrum 
and angular distribution}

\begin{figure}

\centerline{
\epsfig{file=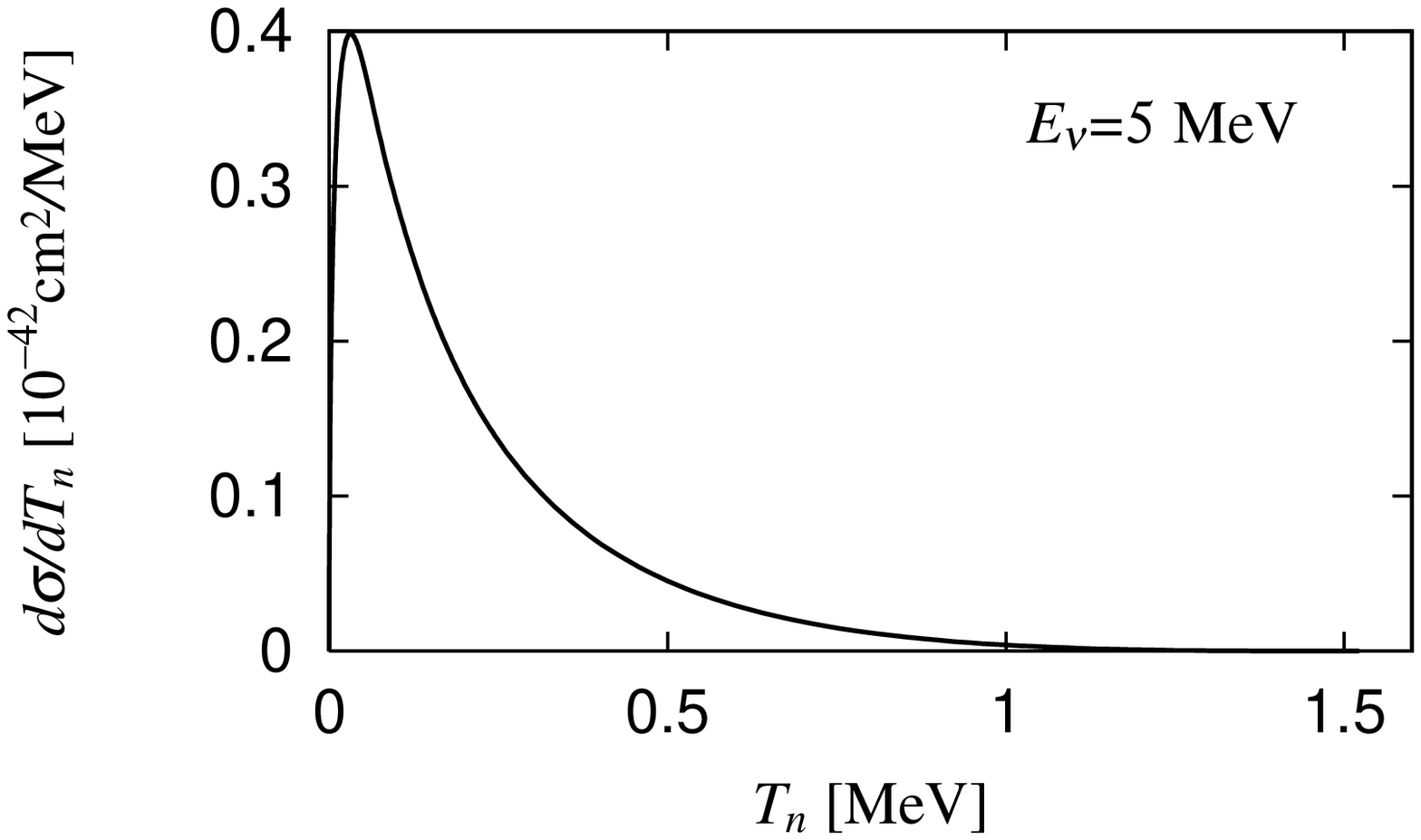,width=9cm}  
}                                     
\centerline{                          
\epsfig{file=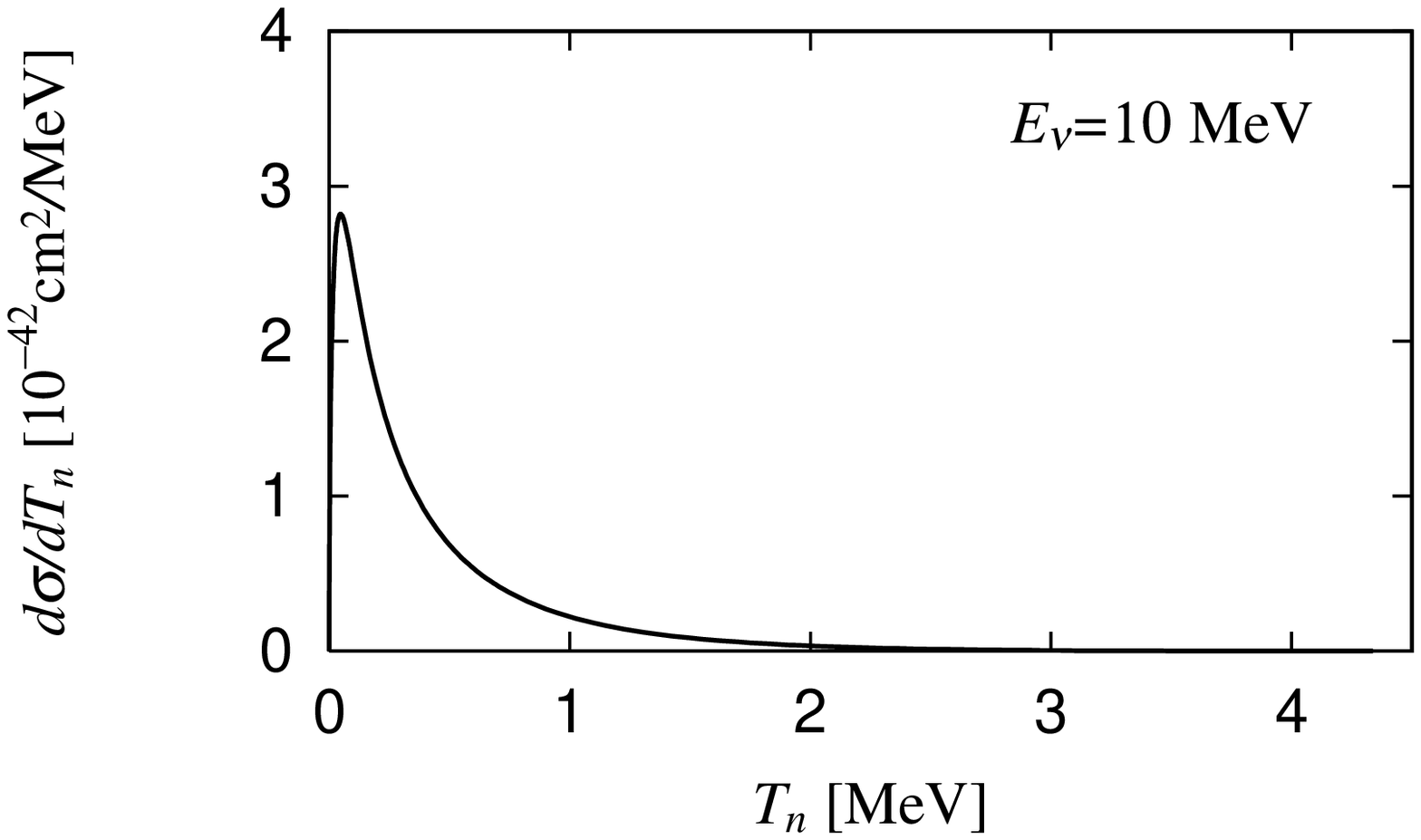,width=9cm} 
}                                     
\centerline{                          
\epsfig{file=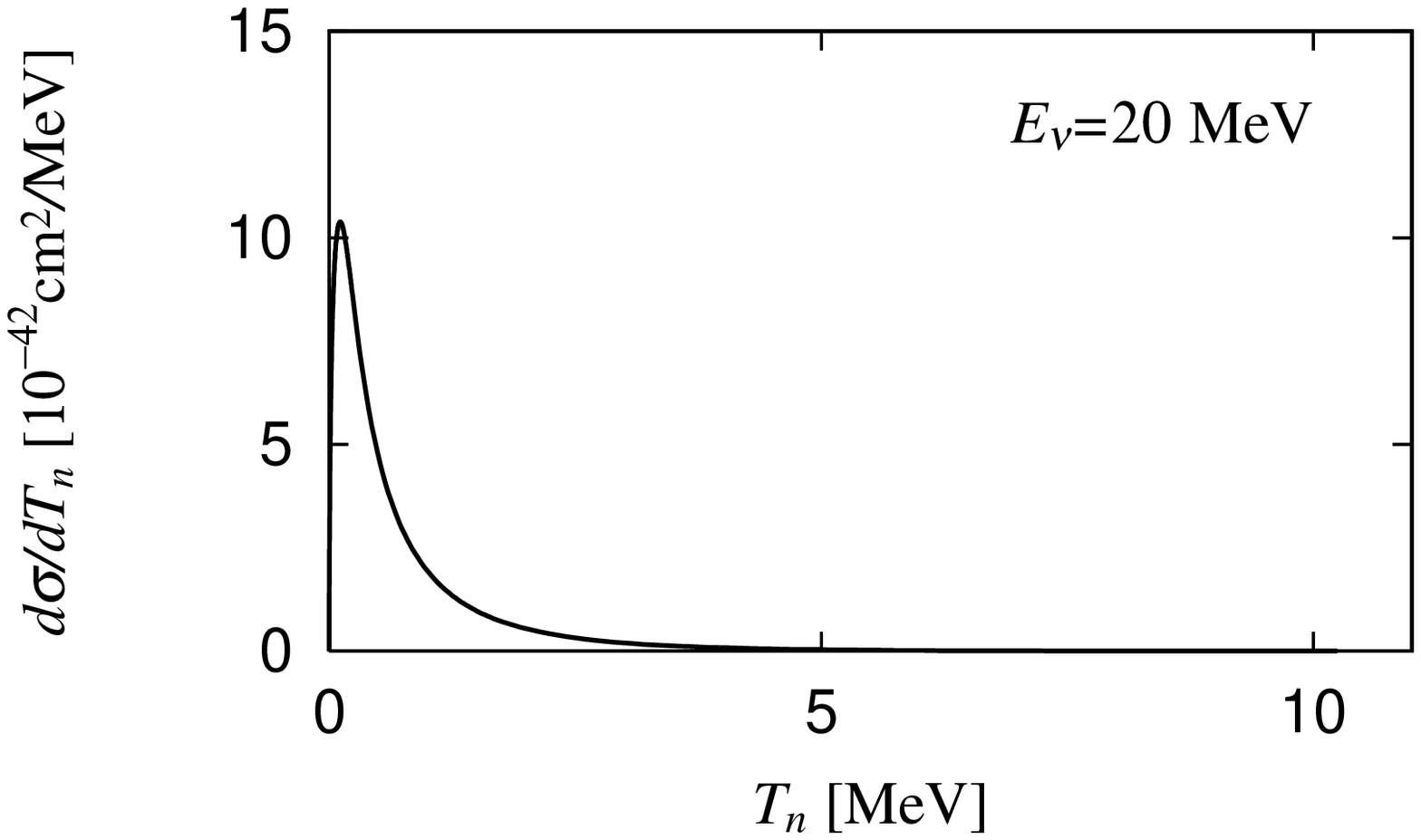,width=9cm} 
}                                     
\centerline{                          
\epsfig{file=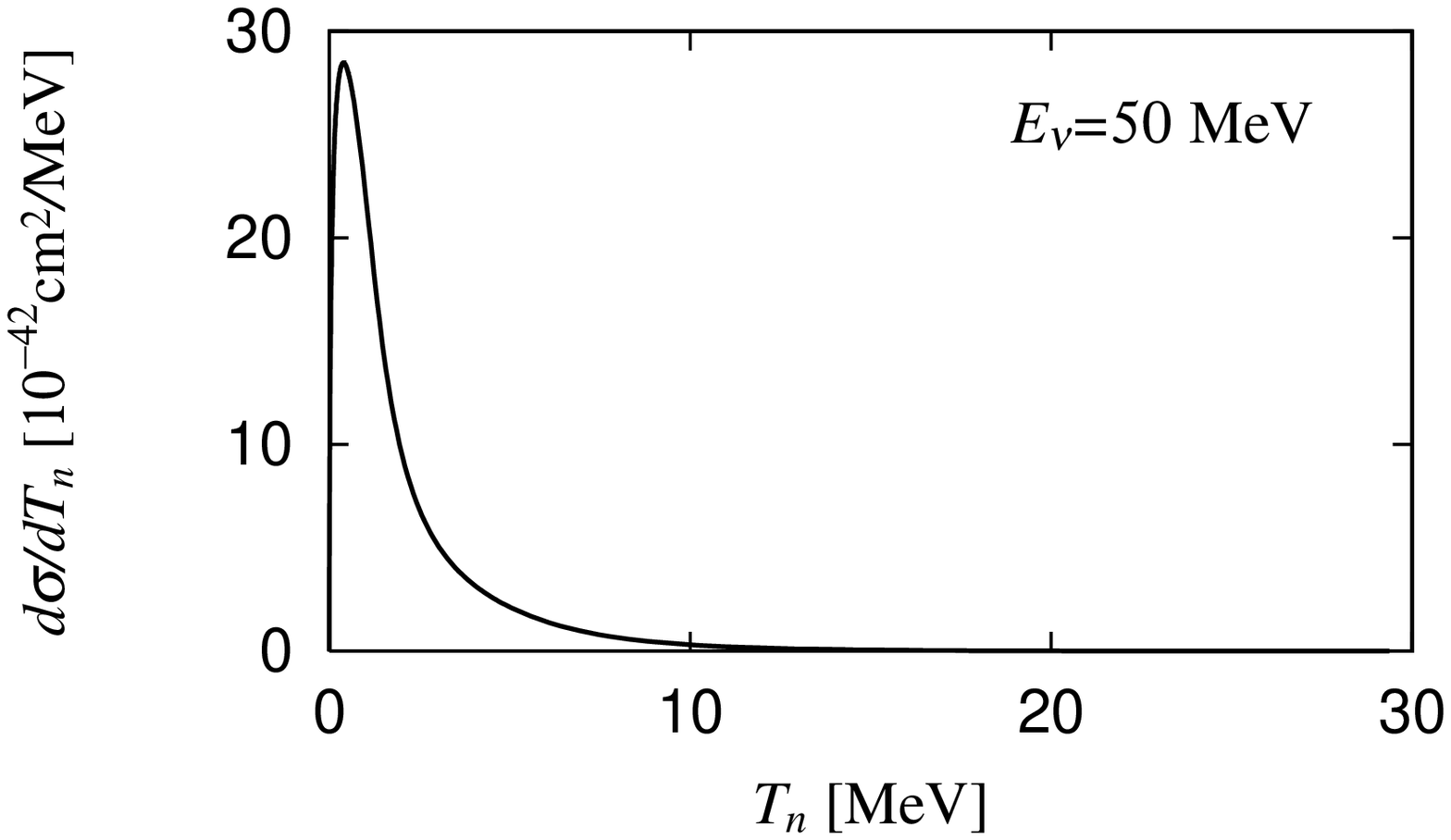,width=9cm} 
}
\caption[]{Neutron energy spectra for
the $\nu d \rightarrow \nu pn$ reaction.}
\label{fig_nspec}
\end{figure}

\begin{figure}

\centerline{
\epsfig{file=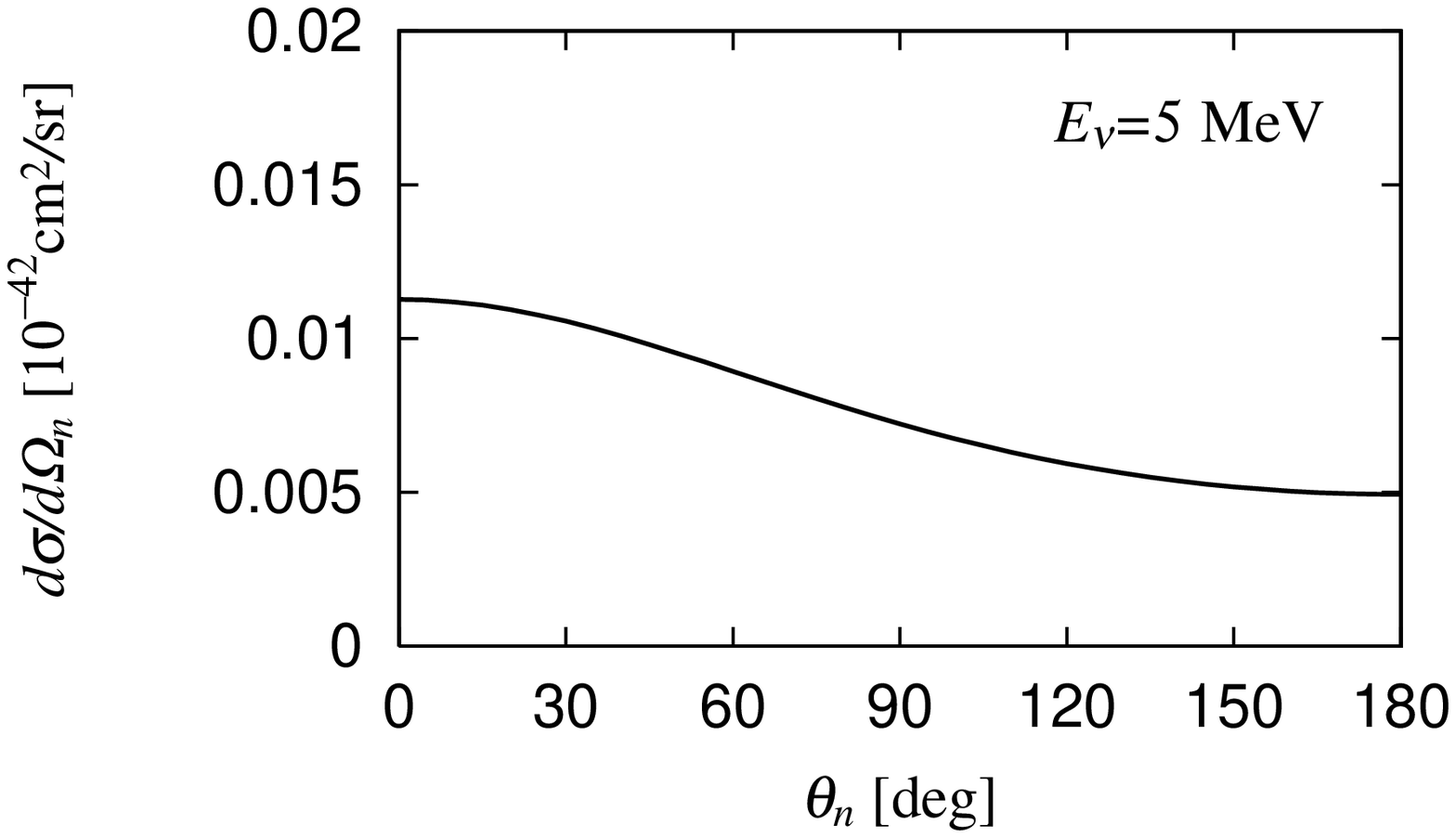,width=9cm}  
}                                      
\centerline{                           
\epsfig{file=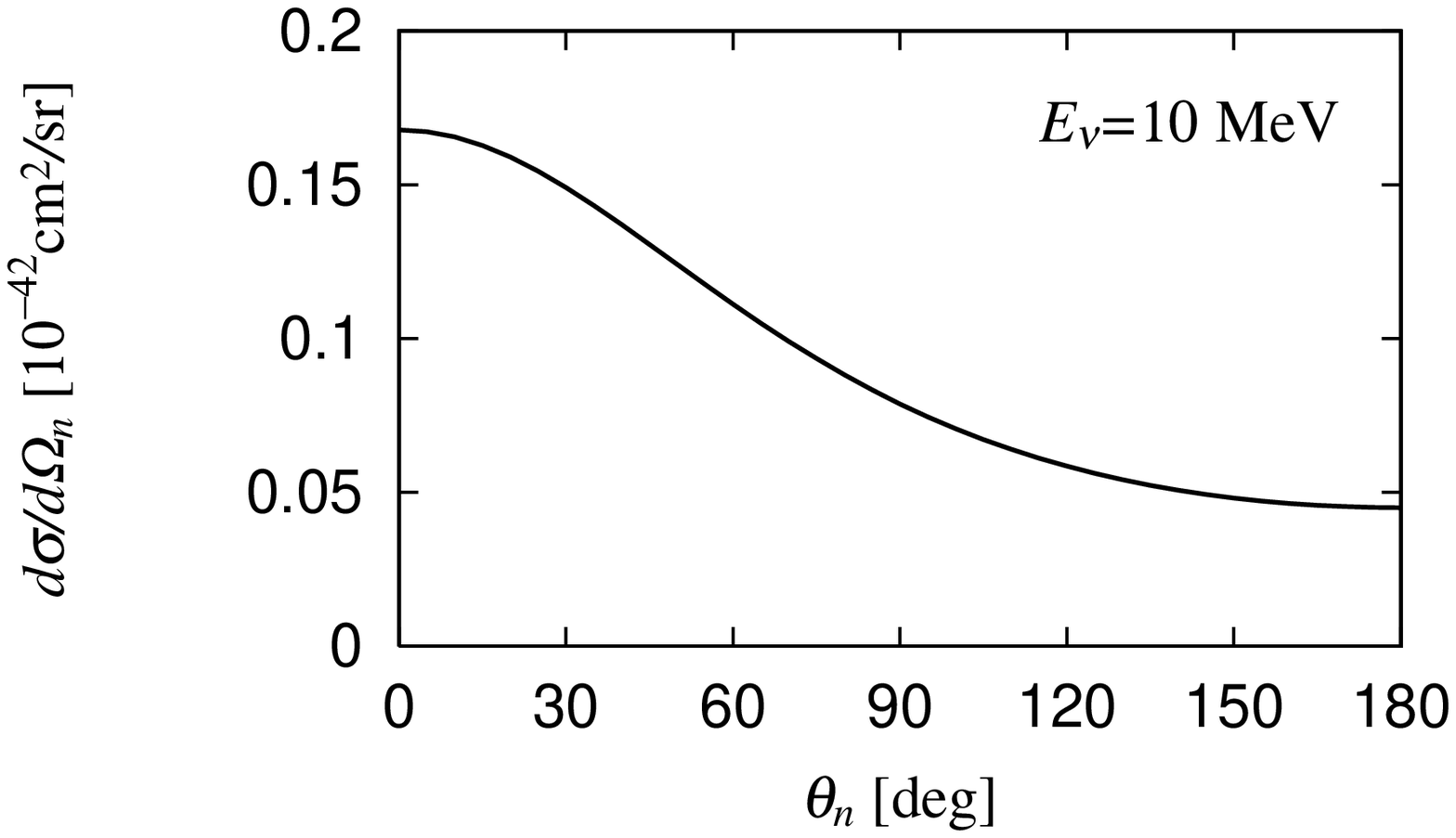,width=9cm} 
}                                      
\centerline{                           
\epsfig{file=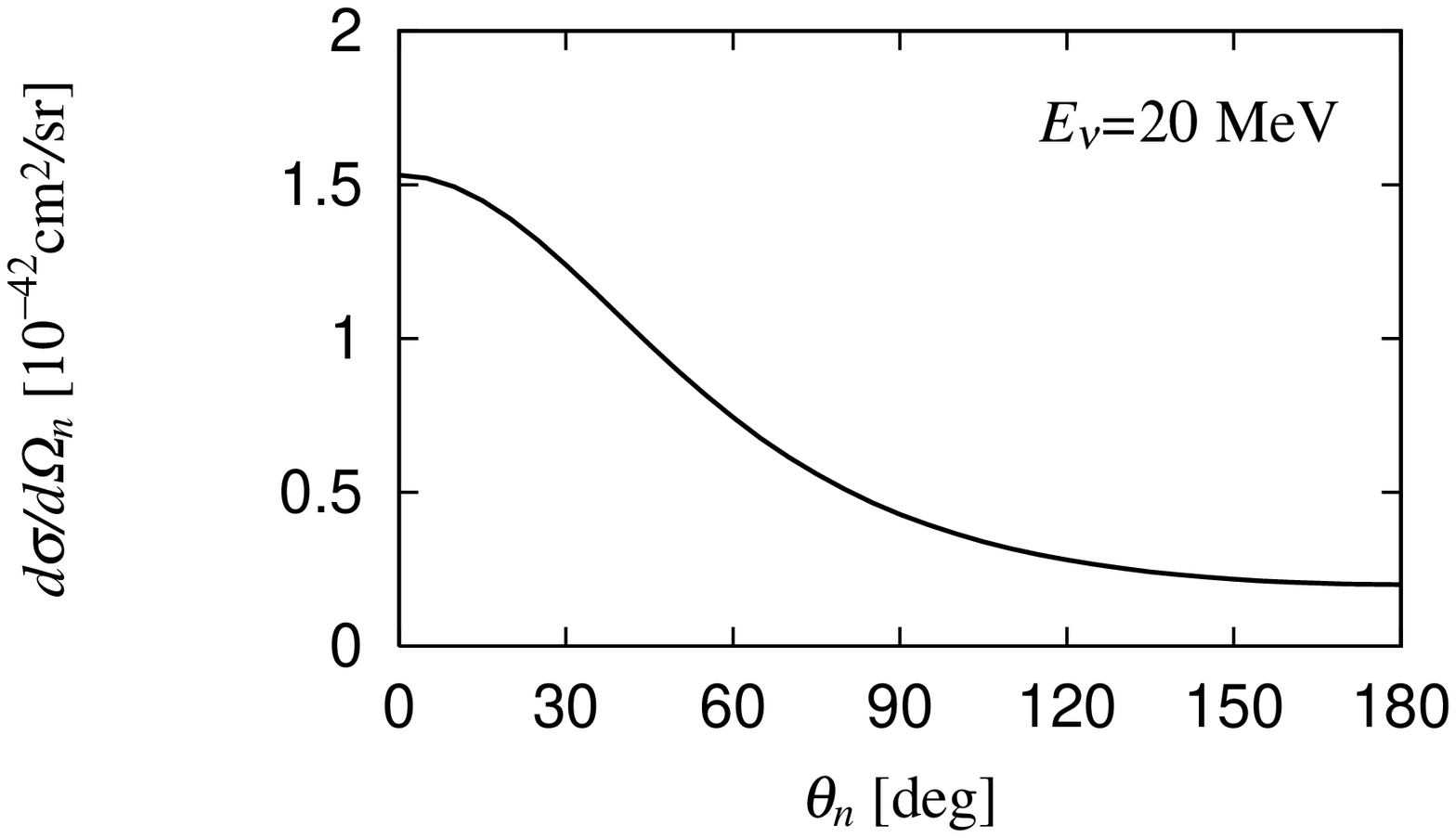,width=9cm} 
}                                      
\centerline{                           
\epsfig{file=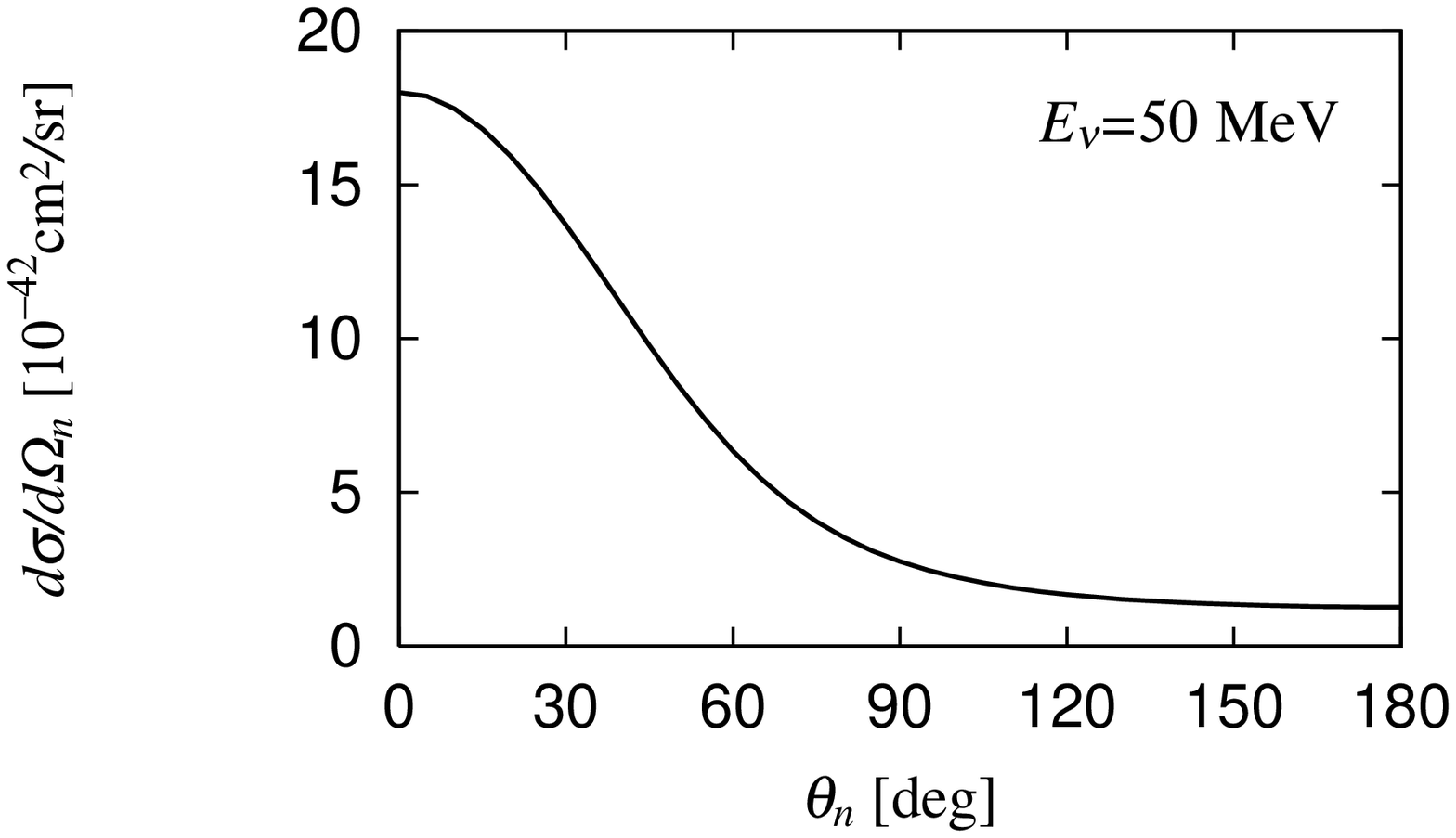,width=9cm}
}
\caption[]{Neutron angular distribution
for the $\nu d \rightarrow \nu pn$ reaction.}
\label{fig_nagl}  
\end{figure}

Finally, we consider the neutron energy spectrum,
$d\sigma/dT_n$
and the neutron angular distribution,
$d\sigma/d\Omega_n$ in Eq.(\ref{eq_cs-nspec}),
for the $\nu + d \rightarrow \nu + p + n$ reaction.
For $E_{\nu} = 5,10,20$, and 50 MeV,
we show $d\sigma/dT_n$ in Fig.\ref{fig_nspec}
and $d\sigma/d\Omega_n$ in Fig.\ref{fig_nagl}.
Once again, we relegate a complete tabulation of 
our numerical results to the web site
mentioned in footnote \ref{foot_web}.
We see from Figs.\ref{fig_nspec} and \ref{fig_nagl}
that the neutron energy spectrum 
has a peak near the lower end, 
and that, unlike the electrons, 
the neutrons are emitted in the forward direction. 
 
\section{Discussion and summary}

Based on a phenomenological Lagrangian approach (PhLA),
we have carried out a detailed study of the $\nu-d$ reactions 
and provided the total cross sections and the differential
cross sections for the electrons and neutrons,
from threshold to $E_\nu=170$ MeV.
We have examined the influence of changes 
in various inputs that feature in our PhLA.
In particular, we have studied to what extent
the use of the modern NN potentials affects the results.
We have also examined the influence of the use of 
the updated input concerning the nucleon weak-interaction
form factors.
The vertex strength that governs the $\Delta$-excitation
axial-vector exchange current has been monitored 
using the photo-reaction.
We have also studied the consequence
of employing the vertex strength
determined with the use of the tritium $\beta$-decay
strength \cite{crsw}.

For the solar energy region, $E_\nu<$ 20 MeV,
the results are summarized as follows.
By comparing our new results with those in the literature,
we have confirmed that the total $\nu-d$ cross sections
are stable within 1\% precision against any changes 
in the input that have been studied,
except for somewhat higher sensitivity
to the strength of the $\Delta$-current
(see below).
The same stability should also exist for the differential
cross sections described in this paper.
The MEX axial-vector current 
in our {\it standard run} 
increases the total cross sections
by $\sim$5\% from the IA values;
we have used the $np \rightarrow \gamma d$
reaction to monitor the dominant part of our MEX current.
Meanwhile, Carlson {\it et al.} \cite{crsw},
in estimating the solar $pp$-fusion cross section,
used the tritium $\beta$-decay lifetime to monitor
a vertex strength that features in the Gamow-Teller
exchange current.  
The results of \cite{crsw} indicates that  
adjusting the MEX strength using the tritium $\beta$-decay rate
could lead to a somewhat reduced MEX amplitude.
If we use the $\Delta$-excitation axial current
renormalized by the tritium $\beta$-decay \cite{crsw},
the MEX current correction to the IA term, 
$[\sigma({\rm IA+MEX})-\sigma(\rm IA)]/\sigma(\rm IA)$,
turns out to be $\sim$2\%, 
instead of 5\% as in our {\it standard run};
see the column labeled $\Delta$(CRSW) 
in TABLE \ref{table_model}.
The difference between our {\it standard run} 
and $\Delta$(CRSW) represents the range of uncertainty
in the present PhLA calculation.
We therefore consider it reasonable to use, 
as the best estimates
of the low-energy $\nu d$ cross sections,
the values given by our {\it standard run} and
attach to them a possible {\it overall} reduction factor $\kappa$, 
with $\kappa$ ranging from 0.96 to 1.
In this language, the ``1$\sigma$" uncertainty
adopted by Bahcall {\it et al.}\cite{bks00}
corresponds to $\kappa=0.95-1$,
which represents the difference between the cross sections
given in YHH \cite{YHH2} and KN \cite{kubo1}.
We have shown that in the ratio 
$R\equiv\sigma({\rm NC})/\sigma({\rm CC})$
the model dependence is reduced down 
to the 1\% level (see TABLE \ref{table_ratio_model}).

At higher incident neutrino energies, 
the results obtained in our {\it standard run}
are somewhat larger than those of KN, 
and the difference reaches $\sim$6\%
towards $E_\nu$=150 MeV.
This difference is caused largely 
by the updated value for the axial-vector mass.
The effect of relativistic kinematics, as discussed here,
has $\sim$1\% effect on the cross sections.
The contributions of the isoscalar current,
which so far has been totally ignored in the literature,
is found to be of 1\% even at $E_\nu\simeq$ 150 MeV.
The importance of the MEX currents relative to the IA
contributions increases monotonically as $E_\nu$ augments.
Towards $E_\nu$=150 MeV, the MEX to IA ratio,
$[\sigma({\rm IA+MEX})-\sigma(\rm IA)]/\sigma(\rm IA)$,
reaches $\sim$8\% 
in our {\it standard run} while this ratio is $\sim$4\%
in the case of $\Delta$(CRSW). 

As mentioned earlier in the text,
the numerical results of this work are
fully documented in tabular or graphical form
at the website referred to in footnote \ref{foot_web}.
It is hoped that those tables and graphs 
are of value for the ongoing 
and future neutrino experiments
that involve deuteron targets.

\begin{center}
{\small\bf{ACKNOWLEDGEMENTS}}\\
\end{center}
The authors wish to express their sincere thanks 
to John Bahcall and Arthur McDonald 
for their interest in the present work.
KK is deeply indebted to Malcolm Butler and Jiunn-Wei Chen
for the enlightening communication on Refs. \cite{EFT,EFT2}.
TS would like to gratefully acknowledge
the hospitality extended to him at the USC Nuclear Physics Group,
where the present collaboration started.
This work is supported in part by
the Ministry of Education, Science, Sport and Culture,
Grant-in-Aid for Scientific Research on Priority Areas (A)(2) 12047233,
and by the National Science Foundation, USA, 
Grant No. PHY-9900756.

\newpage
\appendix
\section{Phase Space Integral and Kinematics} \label{section_kin}

We briefly explain the derivation of the cross section formula
Eq.(\ref{eq_cs}) starting from Eq.(\ref{eq_cs-CC}). The phase space
integral in Eq.(\ref{eq_cs}) is
\begin{eqnarray}
  I &=& \delta^4(k+P-k'-P')\ 
  d\bbox{p'}_1d\bbox{p'}_2d\bbox{k'} \nonumber\\
  &=& \delta(E_{\nu}+M_d-E'_{\ell}
  -\sqrt{\bbox{P'}^2+P'^2_{\mu}}\ )\
  d\bbox{p'}_Ld\bbox{k'},
\end{eqnarray}
where $\bbox{p'}_L=(\bbox{p'}_1-\bbox{p'}_2)/2$ and
$\bbox{P'}=\bbox{q}=\bbox{k}-\bbox{k'}$.

The scattering energy of the final NN distorted wave is given by their
center-of-mass energy $W_{NN}=\sqrt{P_{\mu}^2}$.
The relative momentum in the center-of-mass system,
${p'}^{\mu}$, is given by Lorenz-transforming 
the relative momentum in the lab system as 
\cite{PRD35_3840}
\begin{eqnarray}
{p'}^{\mu} & = & \Lambda^{\mu}_{\nu}\  {p'}_L^{\nu}.
\end{eqnarray}
The magnitude of
$\bbox{p'}$ is related to $W_{NN}$ as 
\begin{eqnarray}
 W_{NN} &=& \sqrt{p'^2 + M^2_{N1}} + \sqrt{p'^2 + M^2_{N2}},
\end{eqnarray}
where $M_{Ni}$ is the mass of the $i$-th nucleon 
in the final state.
The integral over the momentum $\bbox{p'}_L$ is then replaced
by integration over $\bbox{p'}$, which gives rise to 
a Jacobian \cite{PRD35_3840},
\begin{eqnarray}
  d \bbox{p'}_L = J d \bbox{p'},
\end{eqnarray}
with
\begin{eqnarray}
  J = \frac{4 E'_1 E'_2}{W_{NN}(E'_1 + E'_2)},
\end{eqnarray}
where $E'_i$ is the  energy of the $i$-th nucleon 
in the lab system.
Although $J$ depends on the direction of ${\bbox{p'}}$,
we approximate it by
$\bar{J} = \frac{1}{4\pi}\int J
d\Omega_{\bbox{p'}}$;
through a plane-wave calculation,
we have confirmed that this is a good approximation 
in the energy region of our concern.
The phase space integral is then given as
\begin{eqnarray}
  I &=& \delta(E_{\nu}+M_d-E'_{\ell}
  -\sqrt{\bbox{P'}^2+P'^2_{\mu}}\ )\
  \bar{J}d\bbox{p'} d\bbox{k'},
\end{eqnarray}
which leads to Eq.(\ref{eq_cs}).

The kinematically allowed domain of the integral $d\bbox{p'}
d\bbox{k'}$ is determined by a standard procedure. 
We give here the results for the electron energy spectrum,
Eq.(\ref{eq_cs-lepspec}),
for the $\nu_e + d \rightarrow e^-+ p + p$ reaction.
The threshold neutrino energy $E_{\nu}^{th}$ 
for this reaction is given by
\begin{eqnarray}
E_{\nu}^{th} & = & \frac{(2M_p+M_d+m_e)(2M_p-M_d+m_e)}{2M_d}.
\end{eqnarray}
We may specify the allowed region of the electron
energy $E'_e$ by giving the conditions
on the electron momentum $k'$; 
these conditions are
\begin{equation}
  \left\{\begin{array}{@{\,}ll}
  \quad 0 \leq k' \leq k'_{+}& 
  \qquad {\rm for}\quad E_{\nu} \geq E_{\nu}^c \\
\quad k'_{-} \leq k' \leq k'_{+}& 
\qquad {\rm for}\quad E_{\nu}^c \geq
E_{\nu} \geq E_{\nu}^{th},
\end{array} 
\right.  
\end{equation}
where
\begin{eqnarray}
E_{\nu}^c    &\equiv & 
\frac{(2M_p+M_d-m_e)(2M_p-M_d+m_e)}{2(M_d-m_e)},
\end{eqnarray}
and
\begin{eqnarray}
k'_{\pm} = \frac{E_{\nu}X \pm
  (E_{\nu}+M_d)\sqrt{{X}^2-4m_e^2W^2}}{2W^2},
\label{eq_kpm}
\end{eqnarray}
with $W^2=(P + k)^2_{\mu}$ and 
$X \equiv M_d^2 + 2E_{\nu}M_d -4M_p^2 +m_e^2$. 
For given values of $E_{\nu}$ and $E'_e$,
the electron scattering angle $\theta_L$ is restricted as
\begin{eqnarray}
 \mbox{max} \left\{ -1, \frac{2 E'_e(M_d + E_{\nu}) -
  X}{2E_{\nu}k'} \right\} \leq \cos \theta_L \leq 1,
\end{eqnarray}
and the NN scattering energy is specified by $p'$
given as
\begin{eqnarray}
  p' &=&\frac{1}{2}\sqrt{X +
    2E_{\nu}k' \cos \theta_L -2 E'_e(M_d + E_{\nu})}.
\end{eqnarray}
For $E_{\nu} = 150$ MeV, the allowed ranges 
of $\cos \theta_L$ and $p'$
are plotted as functions of $E'_e$
in Fig.7 and Fig.8, respectively;
the dotted area in each figure represents 
the allowed region.
At $E'_e = E'^c_e$, the constraint on $\theta_L$ sets in and
the minimum value of $p'$ becomes zero. 
$E'^c_e$ is determined from the condition:
\begin{eqnarray}
  2 E'^c_e(M_d + E_{\nu}) + 2E_{\nu}k'^c - X = 0.
\end{eqnarray}

\begin{table}
      \footnotesize \tabcolsep=10pt
    \caption[]{Total cross sections for $\nu-d$ reactions in units of
      cm${}^2$. The ``$-x$" in parentheses
      denotes $10^{-x}$; thus an entry like 4.279 (-47)
      stands for 4.279 $\times 10^{-47}$ cm${}^2$.\\}
{   \def\arraystretch{0.6}
    \begin{tabular}{ccccc}
$E_{\nu}$ [MeV]&$\nu\ d \rightarrow \nu\ p\ n$
&$\bar{\nu}\ d \rightarrow\bar{\nu}\ p\ n$
&$\nu_e\  d \rightarrow e^-\ p\ p$
&$\bar{\nu}_e\ d\rightarrow e^+\ n\ n$\\ \hline
 2.0&0.000 (  0)&0.000 (  0)&3.603 (-45)&0.000 (  0)\\
 2.2&0.000 (  0)&0.000 (  0)&7.833 (-45)&0.000 (  0)\\
 2.4&4.279 (-47)&4.248 (-47)&1.404 (-44)&0.000 (  0)\\
 2.6&4.258 (-46)&4.222 (-46)&2.242 (-44)&0.000 (  0)\\
 2.8&1.457 (-45)&1.443 (-45)&3.315 (-44)&0.000 (  0)\\
 3.0&3.355 (-45)&3.320 (-45)&4.639 (-44)&0.000 (  0)\\
 3.2&6.286 (-45)&6.213 (-45)&6.228 (-44)&0.000 (  0)\\
 3.4&1.038 (-44)&1.025 (-44)&8.095 (-44)&0.000 (  0)\\
 3.6&1.574 (-44)&1.553 (-44)&1.025 (-43)&0.000 (  0)\\
 3.8&2.246 (-44)&2.213 (-44)&1.271 (-43)&0.000 (  0)\\
 4.0&3.060 (-44)&3.012 (-44)&1.547 (-43)&0.000 (  0)\\
 4.2&4.024 (-44)&3.956 (-44)&1.855 (-43)&1.115 (-45)\\
 4.4&5.142 (-44)&5.049 (-44)&2.196 (-43)&4.554 (-45)\\
 4.6&6.420 (-44)&6.297 (-44)&2.570 (-43)&1.010 (-44)\\
 4.8&7.860 (-44)&7.702 (-44)&2.978 (-43)&1.787 (-44)\\
 5.0&9.468 (-44)&9.267 (-44)&3.420 (-43)&2.799 (-44)\\
 5.2&1.125 (-43)&1.100 (-43)&3.897 (-43)&4.059 (-44)\\
 5.4&1.320 (-43)&1.289 (-43)&4.410 (-43)&5.578 (-44)\\
 5.6&1.533 (-43)&1.495 (-43)&4.959 (-43)&7.364 (-44)\\
 5.8&1.763 (-43)&1.718 (-43)&5.544 (-43)&9.427 (-44)\\
 6.0&2.012 (-43)&1.958 (-43)&6.166 (-43)&1.177 (-43)\\
 6.2&2.279 (-43)&2.215 (-43)&6.825 (-43)&1.441 (-43)\\
 6.4&2.564 (-43)&2.490 (-43)&7.522 (-43)&1.733 (-43)\\
 6.6&2.868 (-43)&2.782 (-43)&8.258 (-43)&2.056 (-43)\\
 6.8&3.191 (-43)&3.092 (-43)&9.031 (-43)&2.409 (-43)\\
 7.0&3.532 (-43)&3.419 (-43)&9.843 (-43)&2.792 (-43)\\
 7.2&3.893 (-43)&3.764 (-43)&1.069 (-42)&3.206 (-43)\\
 7.4&4.273 (-43)&4.126 (-43)&1.159 (-42)&3.652 (-43)\\
 7.6&4.672 (-43)&4.506 (-43)&1.252 (-42)&4.127 (-43)\\
 7.8&5.091 (-43)&4.904 (-43)&1.349 (-42)&4.635 (-43)\\
 8.0&5.529 (-43)&5.320 (-43)&1.450 (-42)&5.175 (-43)\\
 8.2&5.987 (-43)&5.754 (-43)&1.555 (-42)&5.746 (-43)\\
 8.4&6.464 (-43)&6.206 (-43)&1.664 (-42)&6.349 (-43)\\
 8.6&6.961 (-43)&6.676 (-43)&1.777 (-42)&6.984 (-43)\\
 8.8&7.479 (-43)&7.163 (-43)&1.894 (-42)&7.652 (-43)\\
 9.0&8.016 (-43)&7.669 (-43)&2.016 (-42)&8.351 (-43)\\
 9.2&8.573 (-43)&8.193 (-43)&2.141 (-42)&9.082 (-43)\\
 9.4&9.150 (-43)&8.735 (-43)&2.271 (-42)&9.846 (-43)\\
\end{tabular}}
\clearpage
\noindent
{\small
TABLE I. (continued)\hspace{0.5cm}
Total cross sections for $\nu-d$ reactions in units of
      cm${}^2$. The ``$-x$" in parentheses
      denotes $10^{-x}$; thus an entry like 4.279 (-47)
      stands for 4.279 $\times 10^{-47}$ cm${}^2$.}
{   \def\arraystretch{0.6}
    \begin{tabular}{ccccc}
$E_{\nu}$ [MeV]&$\nu\ d \rightarrow \nu\ p\ n$
&$\bar{\nu}\ d \rightarrow\bar{\nu}\ p\ n$
&$\nu_e\  d \rightarrow e^-\ p\ p$
&$\bar{\nu}_e\ d\rightarrow e^+\ n\ n$\\ \hline
 9.6&9.747 (-43)&9.294 (-43)&2.405 (-42)&1.064 (-42)\\
 9.8&1.036 (-42)&9.872 (-43)&2.544 (-42)&1.147 (-42)\\
10.0&1.100 (-42)&1.047 (-42)&2.686 (-42)&1.233 (-42)\\
10.2&1.166 (-42)&1.108 (-42)&2.833 (-42)&1.322 (-42)\\
10.4&1.234 (-42)&1.171 (-42)&2.984 (-42)&1.415 (-42)\\
10.6&1.304 (-42)&1.236 (-42)&3.139 (-42)&1.510 (-42)\\
10.8&1.376 (-42)&1.303 (-42)&3.299 (-42)&1.609 (-42)\\
11.0&1.450 (-42)&1.372 (-42)&3.463 (-42)&1.712 (-42)\\
11.2&1.526 (-42)&1.442 (-42)&3.631 (-42)&1.817 (-42)\\
11.4&1.604 (-42)&1.514 (-42)&3.804 (-42)&1.925 (-42)\\
11.6&1.684 (-42)&1.588 (-42)&3.981 (-42)&2.037 (-42)\\
11.8&1.767 (-42)&1.664 (-42)&4.163 (-42)&2.152 (-42)\\
12.0&1.851 (-42)&1.741 (-42)&4.349 (-42)&2.270 (-42)\\
12.2&1.938 (-42)&1.821 (-42)&4.539 (-42)&2.392 (-42)\\
12.4&2.026 (-42)&1.902 (-42)&4.734 (-42)&2.516 (-42)\\
12.6&2.117 (-42)&1.985 (-42)&4.933 (-42)&2.644 (-42)\\
12.8&2.210 (-42)&2.069 (-42)&5.137 (-42)&2.775 (-42)\\
13.0&2.305 (-42)&2.156 (-42)&5.346 (-42)&2.909 (-42)\\
13.5&2.551 (-42)&2.379 (-42)&5.887 (-42)&3.258 (-42)\\
14.0&2.811 (-42)&2.614 (-42)&6.456 (-42)&3.626 (-42)\\
14.5&3.084 (-42)&2.860 (-42)&7.054 (-42)&4.015 (-42)\\
15.0&3.371 (-42)&3.117 (-42)&7.681 (-42)&4.422 (-42)\\
15.5&3.671 (-42)&3.385 (-42)&8.338 (-42)&4.849 (-42)\\
16.0&3.984 (-42)&3.663 (-42)&9.024 (-42)&5.295 (-42)\\
16.5&4.311 (-42)&3.953 (-42)&9.740 (-42)&5.760 (-42)\\
17.0&4.651 (-42)&4.253 (-42)&1.049 (-41)&6.244 (-42)\\
17.5&5.006 (-42)&4.564 (-42)&1.126 (-41)&6.747 (-42)\\
18.0&5.374 (-42)&4.886 (-42)&1.207 (-41)&7.268 (-42)\\
18.5&5.755 (-42)&5.218 (-42)&1.291 (-41)&7.809 (-42)\\
19.0&6.151 (-42)&5.561 (-42)&1.378 (-41)&8.367 (-42)\\
19.5&6.560 (-42)&5.915 (-42)&1.468 (-41)&8.944 (-42)\\
20.0&6.984 (-42)&6.279 (-42)&1.561 (-41)&9.539 (-42)\\
20.5&7.421 (-42)&6.653 (-42)&1.657 (-41)&1.015 (-41)\\
21.0&7.872 (-42)&7.038 (-42)&1.757 (-41)&1.078 (-41)\\
21.5&8.338 (-42)&7.434 (-42)&1.859 (-41)&1.143 (-41)\\
22.0&8.817 (-42)&7.839 (-42)&1.965 (-41)&1.210 (-41)\\
22.5&9.311 (-42)&8.255 (-42)&2.074 (-41)&1.278 (-41)\\
23.0&9.819 (-42)&8.681 (-42)&2.187 (-41)&1.348 (-41)\\
23.5&1.034 (-41)&9.117 (-42)&2.303 (-41)&1.420 (-41)\\
24.0&1.088 (-41)&9.564 (-42)&2.422 (-41)&1.494 (-41)\\
24.5&1.143 (-41)&1.002 (-41)&2.545 (-41)&1.569 (-41)\\
\end{tabular}}
\clearpage
\noindent
{\small
TABLE I. (continued)\hspace{0.5cm}
Total cross sections for $\nu-d$ reactions in units of
      cm${}^2$. The ``$-x$" in parentheses
      denotes $10^{-x}$; thus an entry like 4.279 (-47)
      stands for 4.279 $\times 10^{-47}$ cm${}^2$.}
{   \def\arraystretch{0.6}
    \begin{tabular}{ccccc}
$E_{\nu}$ [MeV]&$\nu\ d \rightarrow \nu\ p\ n$
&$\bar{\nu}\ d \rightarrow\bar{\nu}\ p\ n$
&$\nu_e\  d \rightarrow e^-\ p\ p$
&$\bar{\nu}_e\ d\rightarrow e^+\ n\ n$\\ \hline
 25&1.199 (-41)&1.049 (-41)&2.671 (-41)&1.646 (-41)\\
 26&1.317 (-41)&1.145 (-41)&2.933 (-41)&1.805 (-41)\\
 27&1.440 (-41)&1.245 (-41)&3.209 (-41)&1.971 (-41)\\
 28&1.569 (-41)&1.350 (-41)&3.499 (-41)&2.143 (-41)\\
 29&1.704 (-41)&1.458 (-41)&3.803 (-41)&2.322 (-41)\\
 30&1.845 (-41)&1.570 (-41)&4.121 (-41)&2.507 (-41)\\
 31&1.992 (-41)&1.685 (-41)&4.454 (-41)&2.698 (-41)\\
 32&2.145 (-41)&1.805 (-41)&4.802 (-41)&2.896 (-41)\\
 33&2.304 (-41)&1.928 (-41)&5.164 (-41)&3.099 (-41)\\
 34&2.469 (-41)&2.055 (-41)&5.541 (-41)&3.309 (-41)\\
 35&2.640 (-41)&2.186 (-41)&5.934 (-41)&3.525 (-41)\\
 36&2.817 (-41)&2.320 (-41)&6.342 (-41)&3.746 (-41)\\
 37&3.001 (-41)&2.458 (-41)&6.765 (-41)&3.973 (-41)\\
 38&3.190 (-41)&2.600 (-41)&7.204 (-41)&4.206 (-41)\\
 39&3.386 (-41)&2.745 (-41)&7.659 (-41)&4.445 (-41)\\
 40&3.588 (-41)&2.893 (-41)&8.130 (-41)&4.689 (-41)\\
 41&3.796 (-41)&3.045 (-41)&8.617 (-41)&4.938 (-41)\\
 42&4.011 (-41)&3.200 (-41)&9.120 (-41)&5.193 (-41)\\
 43&4.232 (-41)&3.359 (-41)&9.639 (-41)&5.453 (-41)\\
 44&4.459 (-41)&3.521 (-41)&1.018 (-40)&5.718 (-41)\\
 45&4.692 (-41)&3.686 (-41)&1.073 (-40)&5.988 (-41)\\
 46&4.932 (-41)&3.854 (-41)&1.130 (-40)&6.264 (-41)\\
 47&5.178 (-41)&4.026 (-41)&1.188 (-40)&6.544 (-41)\\
 48&5.430 (-41)&4.201 (-41)&1.248 (-40)&6.829 (-41)\\
 49&5.689 (-41)&4.379 (-41)&1.310 (-40)&7.119 (-41)\\
 50&5.954 (-41)&4.559 (-41)&1.374 (-40)&7.413 (-41)\\
 51&6.226 (-41)&4.743 (-41)&1.440 (-40)&7.712 (-41)\\
 52&6.504 (-41)&4.930 (-41)&1.507 (-40)&8.016 (-41)\\
 53&6.788 (-41)&5.120 (-41)&1.575 (-40)&8.324 (-41)\\
 54&7.079 (-41)&5.313 (-41)&1.646 (-40)&8.636 (-41)\\
 55&7.376 (-41)&5.509 (-41)&1.718 (-40)&8.953 (-41)\\
 60&8.957 (-41)&6.528 (-41)&2.107 (-40)&1.060 (-40)\\
 65&1.070 (-40)&7.612 (-41)&2.540 (-40)&1.233 (-40)\\
 70&1.260 (-40)&8.757 (-41)&3.018 (-40)&1.415 (-40)\\
 75&1.465 (-40)&9.959 (-41)&3.540 (-40)&1.606 (-40)\\
 80&1.686 (-40)&1.121 (-40)&4.108 (-40)&1.802 (-40)\\
 85&1.922 (-40)&1.250 (-40)&4.721 (-40)&2.004 (-40)\\
 90&2.172 (-40)&1.383 (-40)&5.378 (-40)&2.212 (-40)\\
 95&2.437 (-40)&1.520 (-40)&6.079 (-40)&2.424 (-40)\\
100&2.715 (-40)&1.660 (-40)&6.824 (-40)&2.640 (-40)\\
105&3.007 (-40)&1.803 (-40)&7.612 (-40)&2.859 (-40)\\
\end{tabular}}
\clearpage
\noindent
{\small
TABLE I. (continued)\hspace{0.5cm}
Total cross sections for $\nu-d$ reactions in units of
      cm${}^2$. The ``$-x$" in parentheses
      denotes $10^{-x}$; thus an entry like 4.279 (-47)
      stands for 4.279 $\times 10^{-47}$ cm${}^2$.}
{   \def\arraystretch{0.6}
    \begin{tabular}{ccccc}
$E_{\nu}$ [MeV]&$\nu\ d \rightarrow \nu\ p\ n$
&$\bar{\nu}\ d \rightarrow\bar{\nu}\ p\ n$
&$\nu_e\  d \rightarrow e^-\ p\ p$
&$\bar{\nu}_e\ d\rightarrow e^+\ n\ n$\\ \hline
110&3.313 (-40)&1.949 (-40)&8.440 (-40)&3.081 (-40)\\
115&3.630 (-40)&2.097 (-40)&9.307 (-40)&3.306 (-40)\\
120&3.958 (-40)&2.247 (-40)&1.021 (-39)&3.532 (-40)\\
125&4.298 (-40)&2.397 (-40)&1.116 (-39)&3.760 (-40)\\
130&4.648 (-40)&2.549 (-40)&1.214 (-39)&3.990 (-40)\\
135&5.009 (-40)&2.702 (-40)&1.315 (-39)&4.220 (-40)\\
140&5.378 (-40)&2.855 (-40)&1.420 (-39)&4.452 (-40)\\
145&5.756 (-40)&3.009 (-40)&1.528 (-39)&4.684 (-40)\\
150&6.143 (-40)&3.163 (-40)&1.639 (-39)&4.918 (-40)\\
155&6.539 (-40)&3.318 (-40)&1.753 (-39)&5.151 (-40)\\
160&6.941 (-40)&3.472 (-40)&1.870 (-39)&5.385 (-40)\\
165&7.350 (-40)&3.627 (-40)&1.989 (-39)&5.621 (-40)\\
170&7.765 (-40)&3.781 (-40)&2.111 (-39)&5.856 (-40)\\
    \end{tabular}}
 \label{table_tot}
\end{table}

\begin{table}
\caption{Contributions of the two lowest partial waves.
For several representative values 
of the incident neutrino energy $E_\nu$
are shown the ratios, 
$\sigma(^1S_0)/\sigma({\rm all})$ and
$\sum_{J=0}^2\sigma(^3P_J)/\sigma({\rm all})$,
as defined in the text.\\}
\begin{tabular}{ccccc}
 & \multicolumn{2}{c}{$d(\nu,\nu)pn$}
 & \multicolumn{2}{c}{$d(\nu,e^-)pp$}\\ \hline
 $E_\nu$[MeV] & $^1S_0$   & $^3P_J$    & $^1S_0$ & $^3P_J$   \\
  5      & 0.999 & 0.001  &  0.999 & 0.001\\
 10      & 0.995 & 0.005  &  0.993 & 0.007\\
 20      & 0.972 & 0.027  &  0.964 & 0.035\\
 50      & 0.827 & 0.158  &  0.804 & 0.182\\
100      & 0.589 & 0.334  &  0.561 & 0.366\\
150      & 0.433 & 0.410  &  0.409 & 0.442
\end{tabular}
\label{table_cnf}
\end{table}

\begin{table}
\caption[]{Contributions of meson exchange currents 
to the total cross section.
The second column (IA) gives the total cross section
obtained with the IA terms alone
(all the cross sections in this table are normalized by
the cross sections obtained in our {\it standard run}). 
The third column ($+\bbox{A}_{MEX}$) shows the cross section
that includes the contribution of 
the space component of the axial exchange current,
while the fourth column ($+A_{KDR,0}$) 
gives the results that contain the additional
contribution of the time component of 
the axial exchange current.
The last column ($+\bbox{V}'_{MEX}$)
gives results including the full exchange currents
using Eq.(\ref{eq_no-Siegert}) for the vector current, 
i.e., without invoking the Siegert theorem. }

\vspace{8mm}
\begin{tabular}{ccccc}
\multicolumn{5}{c}{$d(\nu,\nu)pn$}\\ \hline
$E_\nu$[MeV]&  IA  & $+\bbox{A}_{MEX}$
                       & $+A_{KDR,0}$ & $+\bbox{V}'_{MEX}$ \\
  5 &0.949 & 1.000 &0.999 &1.000  \\
 10 &0.942 & 0.999 &0.999 &1.000  \\
 20 &0.934 & 0.996 &0.996 &1.000  \\
 50 &0.927 & 0.991 &0.991 &0.999  \\
100 &0.925 & 0.984 &0.984 &0.997  \\
150 &0.924 & 0.979 &0.979 &0.996  \\  \hline
\multicolumn{5}{c}{$d(\nu,e^-)pp$}\\ \hline
$E_\nu$[MeV]&  IA  & $+\bbox{A}_{MEx}$
                       & $+A_{KDR,0}$ & $+\bbox{V}'_{MEX}$ \\
  5 &0.952 &0.999 &0.999 &1.000 \\
 10 &0.945 &0.997 &0.997 &1.000 \\
 20 &0.937 &0.994 &0.994 &1.000 \\
 50 &0.928 &0.985 &0.985 &0.999 \\
100 &0.924 &0.974 &0.974 &0.995 \\
150 &0.922 &0.966 &0.966 &0.993 
\end{tabular}
\label{table_crnt}
\end{table}

\begin{table}
\caption{Model dependence of total cross sections.
 The second column (YHH) and the third column (KN)
 give $\sigma({\rm YHH})/\sigma(standard\;run)$
 and $\sigma({\rm KN})/\sigma(standard\;run)$,
 respectively.
 The fourth column (Reid93) [fifth column (NIJ II)]
 gives the ratio of the total cross section obtained 
 with the use of the Reid 93 potential [Nijmegen II potential]
 to that of our {\it standard run}.
 The last column ($\Delta$(CRSW)) gives the ratio
 of the total cross section obtained 
 with the $\Delta$-current of Carlson {\it et al.}
 to that of our {\it standard run}. \\}
\begin{tabular}{cccccc}
 \multicolumn{5}{c}{$d(\nu,\nu)pn$}\\ \hline
$E_\nu$[MeV] &YHH&KN&Reid93&NIJ II&$\Delta({\rm CRSW})$\\
  5& 0.962& 1.002& 0.997& 1.002& 0.965\\
 10& 0.955& 1.003& 0.998& 1.002& 0.961\\
 20& 0.946& 1.000& 0.998& 1.001& 0.956\\
 50& 0.964& 0.993& 0.999& 1.000& 0.953\\
100& 0.961& 0.971& 1.000& 1.000& 0.953\\
150& 0.915& 0.943& 1.000& 0.999& 0.954\\ \hline
 \multicolumn{5}{c}{$d(\nu,e^-)pp$}\\ \hline
$E_\nu$[MeV]&YHH&KN&Reid93&NIJ II&$\Delta({\rm CRSW})$\\
  5& 0.956& 1.019& 1.003& 1.003& 0.968\\
 10& 0.949& 1.008& 1.003& 1.002& 0.964\\
 20& 0.948& 1.002& 1.002& 1.001& 0.959\\
 50& 0.961& 0.990& 1.001& 1.000& 0.956\\
100& 0.955& 0.968& 1.001& 0.999& 0.956\\
150& 0.897& 0.941& 1.001& 0.999& 0.956
\end{tabular}
\label{table_model}
\end{table}

\begin{table}
\caption{Model dependence of 
$R \equiv \sigma (NC)/\sigma (CC)
\equiv \sigma(\nu d \rightarrow \nu np)
/\sigma(\nu_e d\rightarrow e^-pp)$.
For representative values of $E_\nu$,
$R$ for our {\it standard run} is given 
in the second column.
The third through the sixth columns
give $R_a$, with $a$ = IA, Reid93, NIJ II, and $\Delta$(CRSW),
normalized by $R_{standard \; run}$.
See also the caption for TABLE \ref{table_model}.\\}
\begin{tabular}{cccccc}
$E_\nu$[MeV] &$R_{standard \; run}$&IA&Reid93&NIJ II&$\Delta({\rm CRSW})$\\
  5& 0.277& 0.997& 0.994& 0.999& 0.997\\
 10& 0.410& 0.997& 0.996& 1.000& 0.997\\
 20& 0.447& 0.997& 0.997& 1.000& 0.997\\
 50& 0.433& 0.999& 0.998& 1.000& 0.997\\
100& 0.398& 1.001& 0.999& 1.000& 0.997\\
150& 0.375& 1.003& 1.000& 1.001& 0.998
\end{tabular}

\label{table_ratio_model}
\end{table}

\newpage

\newpage
\newpage
\newpage
\newpage
\newpage
\begin{figure}
\centerline{
\epsfig{file=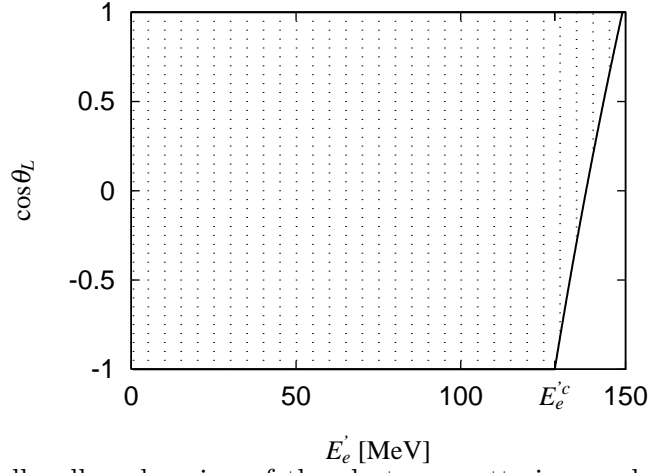,width=9cm}  
}
\label{fig_kin_angle}
\caption[]{Kinematically allowed region of 
the electron scattering angle $\theta_L$
in the $\nu_e d \rightarrow e^- pp$ reaction
at $E_{\nu} = 150$ MeV.
The dotted area represents the allowed region.
The constraint on $\theta_L$ sets in
at $E'_e = E'^c_e$.}
\end{figure}
\begin{figure}
\centerline{                          
\epsfig{file=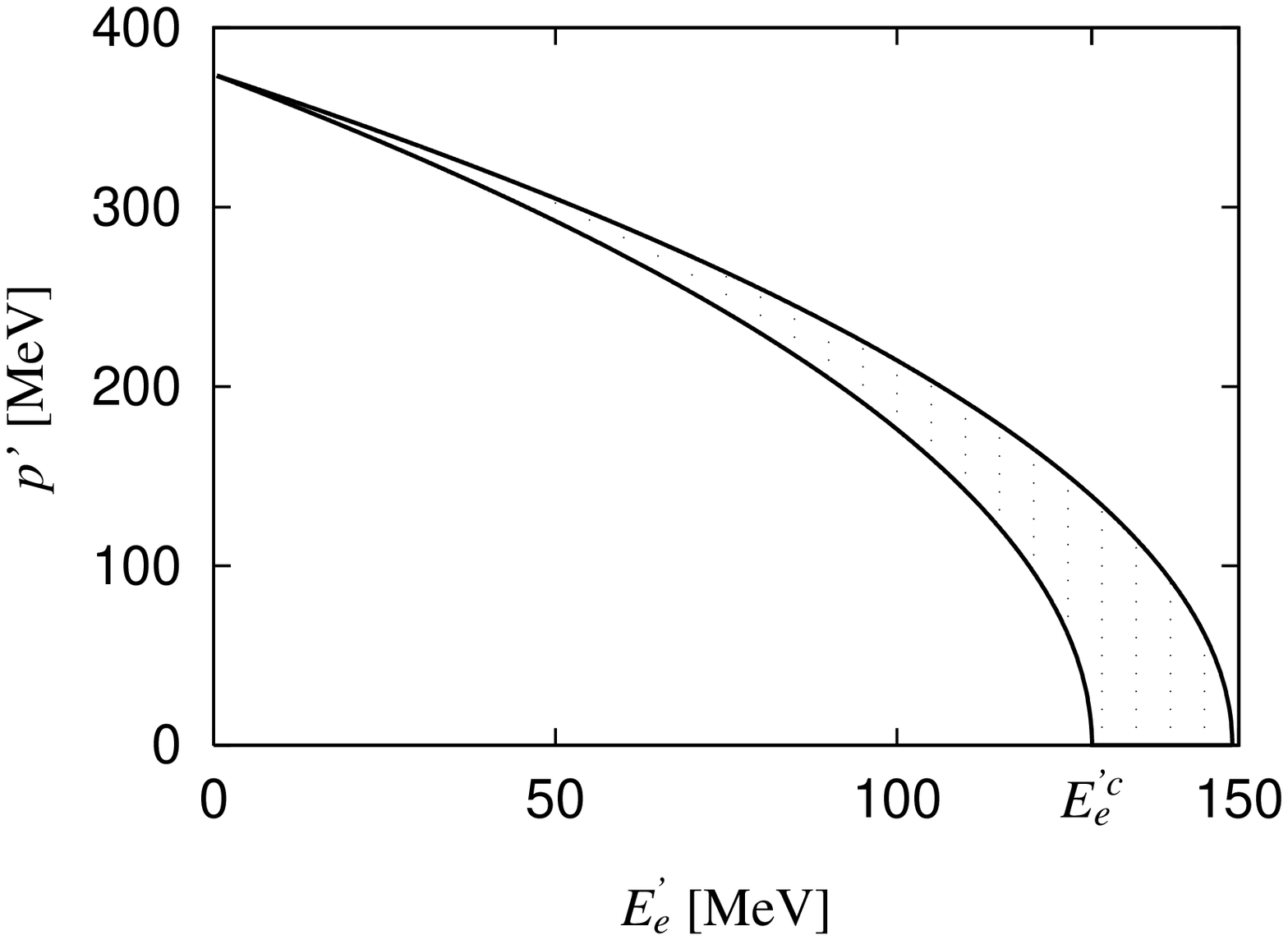,width=9cm} 
}
\label{fig_kin_pc}
\caption[]{Kinematically allowed region of $p'$,
the relative momentum of the final two nucleons
in the $\nu_e d\rightarrow e^- pp$ reaction 
at $E_{\nu} = 150$ MeV. 
The lower limit of $p'$ reaches zero 
at $E'_e = E'^c_e$.}
\end{figure}


\begin{thebibliography}{99}

\bibitem{ellis}
S. D. Ellis and J. N. Bahcall, Nucl. Phys. {\bf A114}, 636 (1968).

\bibitem{PRD12-3673} 
A. Ali and C. A. Dominguez, Phys. Rev. D {\bf 12}, 3673 (1975).

\bibitem{NPA294_473}
H. C. Lee, Nucl. Phys. {\bf A294}, 473 (1978).

\bibitem{avi} 
F. T. Avignone III, Phys. Rev. D {\bf 24}, 778 (1981).

\bibitem{nozetal} 
S. Nozawa, Y. Kohyama, T. Kaneta and K. Kubodera, 
J. Phys. Soc. Jpn. {\bf 55}, 2636 (1986).

\bibitem{bkn} 
J. N. Bahcall, K. Kubodera and S. Nozawa, Phys. Rev. D
{\bf 38}, 1030 (1988).

\bibitem{YHH1}
S. Ying, W. C. Haxton and E. M. Henley,
Phys. Rev. D {\bf 40}, 3211 (1989).

\bibitem{TKK}
N. Tatara, Y. Kohyama and K. Kubodera,
Phys. Rev. C {\bf 42}, 1694 (1990).

\bibitem{DK1}
M. Doi and K. Kubodera, Phys. Rev. C {\bf 45}, 1988 (1992).

\bibitem{YHH2}
S. Ying, W. C. Haxton and E. M. Henley,
Phys. Rev. C {\bf 45}, 1982 (1992).

\bibitem{kk92}
Y. Kohyama and K. Kubodera, 
USC(NT)-Report-92-1 (1992) (unpublished);
M. Doi and K. Kubodera (unpublished).

\bibitem{kubo1}
K. Kubodera and S. Nozawa,
Int. J. Mod. Phys. E {\bf 3}, 101 (1994).

\bibitem{EFT} 
M. Butler and J.-W. Chen, Nucl. Phys. {\bf A675}, 
575 (2000).

\bibitem{EFT2} 
M. Butler, J.-W. Chen and X. Kong, nucl-th/0008032.

\bibitem{SNO}
The SNO Collaboration,
Phys. Lett. {\bf B194}, 321 (1987);
nucl-ex/9910016;
G. T. Ewan {\it et al.}, Sudbury Neutrino Observatory
Proposal SNO-87-12, 1987.

\bibitem{cr} 
M. Chemtob and M. Rho, Nucl. Phys. {\bf A163}, 1 (1971).

\bibitem{it77} 
E. Ivanov and E. Truhlik, 
Nucl. Phys. {\bf A316}, 451 (1979); {\bf A316}, 437 (1979).

\bibitem{towner} 
I. S. Towner, Phys. Rep. {\bf 155}, 263 (1987). 

\bibitem{riska-brown}
D. O. Riska and G. E. Brown, Phys. Lett. {\bf 38B}, 193 (1972).

\bibitem{fm}
See e.g., B. Frois and J.-F. Mathiot, Com. Part. Nucl. Phys.
{\bf 18}, 291 (1989), and references therein.

\bibitem{n-cap}
T. Sato, T. Niwa and H. Ohtsubo, 
in {\it Proceedings  of the IV International Symposium on Weak
and Electromagnetic Interactions in Nuclei},
edited by H. Ejiri, T. Kishimoto and T. Sato
(World Scientific, Singapore, 1995), p. 488.

\bibitem{drr}
F. Dautry, M. Rho and D. O. Riska, 
Nucl. Phys. {\bf A264}, 507 (1976). 

\bibitem{Doi-muon} 
M. Doi, T. Sato, H. Ohtsubo and M. Morita, 
Nucl. Phys. {\bf A511}, 507 (1990).

\bibitem{crsw} 
J. Carlson, D. O. Riska, R. Schiavilla and R. B. Wiringa,
Phys. Rev. C {\bf 44}, 619 (1991).

\bibitem{Willis} 
S. E. Willis {\it et al.}, Phys. Rev. Lett. {\bf 44}, 522 (1980).

\bibitem{bugey} 
  S. P. Riley, Z. D. Greenwood, W. R. Kropp, L. R. Price, F. Reines,
  H. W. Sobel, Y. Declais, A. Etenko and M. Skorokhvatov,
  Phys. Rev. C {\bf 59}, 1780 (1999). 

\bibitem{pmr} 
T.-S. Park, D.-P. Min and M. Rho,
Phys. Rev. Lett. {\bf 74}, 4153 (1995);
Nucl. Phys. {\bf A596}, 515 (1996).

\bibitem{pkmr} 
T.-S. Park, K. Kubodera, D.-P. Min and M. Rho,
Phys. Rev. C {\bf 58}, 637 (1998);
Astrophys. J. {\bf 507}, 443 (1998);
Nucl. Phys. {\bf A646}, 83 (1999);
Phys. Lett. {\bf B472}, 232 (2000).

\bibitem{kol} 
U. van Kolck, Prog. Part. Nucl. Phys. {\bf 43}, 337 (1999),
and references therein.

\bibitem{ksw} 
D. B. Kaplan, M. J. Savage and M. B. Wise,
Nucl. Phys. {\bf B478}, 629 (1996); 
Phys. Lett. {\bf B424}, 390 (1998); 
Nucl. Phys. {\bf B534}, 329 (1998); 
Phys. Rev. C {\bf 59}, 617 (1999).

\bibitem{pds2} 
J.-W. Chen, H. W. Grie\ss hammer, M. J. Savage and R. P. Springer,
Nucl. Phys. {\bf A644}, 221 (1998); {\bf A644}, 245 (1998);
D. B. Kaplan, M. J. Savage, R. P. Springer and M. B. Wise, 
Phys. Lett. {\bf B449}, 1 (1999);
T. Mehen and I. W. Stewart, Phys. Lett. {\bf B445}, 378 (1998);
X. Kong and F. Ravndal, Phys. Lett. {\bf B450}, 320 (1999). 


\bibitem{pt99} 
D. R. Phillips and T. D. Cohen, Nucl. Phys. {\bf A668}, 45 (2000), and
references therein. 

\bibitem{bks00} 
J. N. Bahcall, P. I. Krastev and A. Yu. Smirnov,
hep-ph/0002293. 

\bibitem{cvm} 
M. Gourdin, Phys. Rep. {\bf 11}, 29 (1974).

\bibitem{avm} 
T. Kitagaki {\it et al.}, Phys. Rev. D {\bf 42}, 1331 (1990).


\bibitem{oz70}
V. I. Ogievetsky and B. M. Zupnik, 
Nucl. Phys. {\bf B24}, 612 (1970).

\bibitem{KDR} 
K. Kubodera, J. Delorme and M. Rho, 
Phys. Rev. Lett. {\bf 40}, 755 (1978).

\bibitem{krt92}
M. Kirchbach, D. O. Riska and K. Tsushima, 
Nucl. Phys. {\bf A542}, 616 (1992).

\bibitem{tow92}
I. S. Towner, 
Nucl. Phys. {\bf A542}, 631 (1992).

\bibitem{isobar} 
S. L. Adler,
Ann. Phys. {\bf 50}, 189 (1968);
P. Guichon, M. Giffon, J. Joseph, R. Laverri\` ere and C. Samour,
Z. Phys. A {\bf 285}, 183 (1978).

\bibitem{cutoff}
J. W. Durso, A. D. Jackson and B. J. Verwest, 
Nucl. Phys. {\bf A282}, 404 (1977); 
F. Iachello, A. D. Jackson and A. Lande, 
Phys. Lett. {\bf 43B}, 191 (1973).

\bibitem{anlv18} 
R. B. Wiringa, V. G. J. Stoks and R. Schiavilla, 
Phys. Rev. C {\bf 51}, 38 (1995).

\bibitem{nij-Reid93} 
V. G. J. Stoks, R. A. M. Klomp, C. P. F. Terheggen, and J. J. de
Swart, Phys. Rev. C {\bf 49}, 2950 (1994).

\bibitem{walecka} 
J. D. Walecka, in { \it Muon Physics}, 
edited by V. W. Hughes and C. S. Wu
(Academic, New York, 1975), Vol. 2, p. 113.

\bibitem{fermi-fn} 
See e.g., M. Morita, {\it Beta Decay and Muon Capture},
(W. A. Benjamin, Inc., 1973) p. 27.

\bibitem{cox} 
A. E. Cox, S. A. R. Wynchank and C. H. Collie,
Nucl. Phys. {\bf 74}, 497 (1965).

\bibitem{nagai} 
T. S. Suzuki, Y. Nagai, T. Shima, T. Kikuchi, H. Sato, T. Kii and
M. Igashira, Astrophys. J. {\bf 439}, 59 (1995).
  
\bibitem{gamma-d1} 
Y. Birenbaum, S. Kahane and R. Moreh, Phys. Rev. C {\bf 32}, 1825 (1985).

\bibitem{gamma-d2} 
R. Bernabei {\it et al.}, Phys. Rev. Lett. {\bf 57}, 1542 (1986).


\bibitem{MAINZ} 
A. Liesenfeld {\it et al.}, Phys. Lett. {\bf B468}, 20 (1999).

\bibitem{PRD35_3840} 
W. Gl\"ockle and Y. Nogami, Phys. Rev. D {\bf 35}, 3840 (1987).

\end{thebibliography}
\end{document}